\documentclass{aa} 
\usepackage{xcolor}
\usepackage{support-caption}
\usepackage{subcaption}
\usepackage{graphicx}
\usepackage[flushleft]{threeparttable}

\newcommand{\nii}{[\textrm{N}\textsc{ii}]}
\newcommand{\hii}{\textrm{H}\textsc{ii}}

\newcommand{\oiii}{[\textrm{O}\textsc{iii}]}
\newcommand{\oii}{[\textrm{O}\textsc{ii}]}

\newcommand{\oiilam}{[\textrm{O}\textsc{ii}]\ensuremath{\lambda3727}}
\newcommand{\mgii}{[\textrm{Mg}\textsc{ii}]\ensuremath{\lambda\lambda2796,2803}}

\newcommand{\oiiialone}{[\textrm{O}\textsc{iii}]}

\newcommand{\oiiidoub}{[\textrm{O}~\textsc{iii}]\ensuremath{\lambda\lambda4959,5007}}
 
\newcommand{\ha}{\ifmmode {\rm H}\alpha \else H$\alpha$\fi}
\newcommand{\hb}{\ifmmode {\rm H}\beta \else H$\beta$\fi}
\newcommand{\lya}{\ifmmode {\rm Ly}\alpha \else Ly$\alpha$\fi}
\newcommand{\pg}{\ifmmode {\rm P}\gamma \else Pa$\gamma$\fi}
\newcommand{\lyb}{\ifmmode {\rm Ly}\beta \else Ly$\beta$\fi}
\newcommand{\lyg}{\ifmmode {\rm Ly}\gamma \else Ly$\gamma$\fi}

\newcommand{\flyc}{\ifmmode \mathrm{f}_\mathrm{esc}\mathrm{(LyC)} \else $\mathrm{f}_\mathrm{esc}\mathrm{(LyC)}$\fi}

\def\ergs{\ifmmode \mathrm{erg\hspace{1mm}s}^{-1} \else erg s$^{-1}$\fi}

\def\micron{\ifmmode \mu\mathrm{m} \else $\mu$m\fi}
\def\msun{\ifmmode \mathrm{M}_{\odot} \else M$_{\odot}$\fi}
\def\msunyr{\ifmmode \mathrm{M}_{\odot} \hspace{1mm}{\rm yr}^{-1} \else $\mathrm{M}_{\odot}$ yr$^{-1}$\fi}
\def\zsun{\ifmmode Z_{\odot} \else Z$_{\odot}$\fi}
\def\lsun{\ifmmode L_{\odot} \else L$_{\odot}$\fi}
\def\mstar{\ifmmode \mathrm{M}_{\star} \else M$_{\star}$\fi}
\newcommand{\hst}{\textit{HST}}
\newcommand{\jwst}{\textit{JWST}}

\newcommand{\NIRSpec}{\textit{NIRSpec}}
\newcommand{\NIRCam}{\textit{NIRCam}}

\usepackage{txfonts}
\newcommand{\orcid}[1]{\href{https://orcid.org/#1}{\includegraphics[width=10pt]{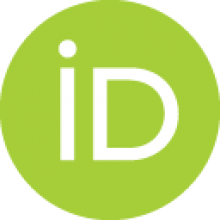}}}
\usepackage{hyperref}
\hypersetup{colorlinks, % = true per default
 linkcolor = [rgb]{1.,0.2,0.2},
 urlcolor = [rgb]{0.7,0.2,0.2},
 citecolor = [rgb]{0.1,0.4,1.}
}
\begin{document} 

\title{Closing in on the sources of cosmic reionization: First results from the GLASS-JWST program}

 \subtitle{}
 \author{S. Mascia \orcid{0000-0002-9572-7813}
 \inst{1,2}
 \and L. Pentericci \orcid{0000-0001-8940-6768}
 \inst{1}
 \and A. Calabrò \orcid{0000-0003-2536-1614}
 \inst{1}
 \and T. Treu \orcid{0000-0002-8460-0390}
 \inst{3}
 \and P. Santini \orcid{0000-0002-9334-8705}
 \inst{1}
 \and L. Yang \orcid{0000-0002-8434-880X}
 \inst{4}
 \and L. Napolitano \orcid{0000-0002-8951-4408}
 \inst{1}
 \and G. Roberts-Borsani \orcid{0000-0002-4140-1367}
 \inst{3}
 \and P. Bergamini \orcid{0000-0003-1383-9414}
 \inst{5,6}
 \and C. Grillo \orcid{0000-0002-5926-7143}
 \inst{5,7}
 \and P. Rosati \orcid{0000-0002-6813-0632}
 \inst{6,8}
 \and B. Vulcani \orcid{0000-0003-0980-1499}
 \inst{9}
 \and M. Castellano \orcid{0000-0001-9875-8263}
 \inst{1}
 \and 
 K. Boyett \orcid{0000-0003-4109-304X}
 \inst{10,11}
 \and A. Fontana \orcid{0000-0003-3820-2823}
 \inst{1}
 \and K. Glazebrook \orcid{0000-0002-3254-9044}
\inst{17}
 \and A. Henry \orcid{0000-0002-6586-4446}
 \inst{12,13}
 \and C. Mason \orcid{0000-0002-3407-1785}
 \inst{14,15}
 \and E. Merlin \orcid{0000-0001-6870-8900}
 \inst{1}
 \and T. Morishita \orcid{0000-0002-8512-1404}
 \inst{16}
 \and T. Nanayakkara \orcid{0000-0003-2804-0648}
 \inst{17}
 \and D. Paris \orcid{0000-0002-7409-8114}
 \inst{1}
 \and N. Roy \orcid{0000-0002-4430-8846}
 \inst{13}
 \and H. Williams \orcid{0000-0002-1681-0767}
 \inst{18}
 \and X. Wang \orcid{0000-0002-9373-3865}
 \inst{19,20,21}
 \and G. Brammer \orcid{0000-0003-2680-005X}
 \inst{14,15}
 \and M. Brada\v{c} \orcid{0000-0001-5984-0395}
\inst{}\  \inst{20,21}
 \and {W. Chen} \orcid{0000-0003-1060-0723}
\inst{18}
 \and P. L. Kelly \orcid{0000-0003-3142-997X}
 \inst{18}
 \and A. M. Koekemoer \orcid{0000-0002-6610-2048}
 \inst{12}
 \and
 M. Trenti \orcid{0000-0001-9391-305X}
 \inst{10,11}
 \and 
 R. A. Windhorst \orcid{0000-0001-8156-6281}
 \inst{19}
 }
 \institute{\textit{INAF – Osservatorio Astronomico di Roma, via Frascati 33, 00078, Monteporzio Catone, Italy}\\
 \email{sara.mascia@inaf.it}
 \and 
 \textit{Dipartimento di Fisica, Università di Roma Tor Vergata,
Via della Ricerca Scientifica, 1, 00133, Roma, Italy}
\and 
\textit{Department of Physics and Astronomy, University of California, Los Angeles, 430 Portola Plaza, Los Angeles, CA 90095, USA}
\and
\textit{Kavli Institute for the Physics and Mathematics of the Universe, The University of Tokyo, Kashiwa, Japan 277-8583}
\and
\textit{Dipartimento di Fisica, Università degli Studi di Milano, Via Celoria 16, I-20133 Milano, Italy}
\and
\textit{INAF - OAS, Osservatorio di Astrofisica e Scienza dello Spazio di Bologna, via Gobetti 93/3, I-40129 Bologna, Italy}
\and 
\textit{INAF - IASF Milano, via A. Corti 12, I-20133 Milano, Italy} 
 \and
 \textit{Dipartimento di Fisica e Scienze della Terra, Università degli Studi di Ferrara, Via Saragat 1, I-44122 Ferrara, Italy}
 \and 
\textit{INAF Osservatorio Astronomico di Padova, vicolo dell'Osservatorio 5, 35122 Padova, Italy}
 \and
 \textit{School of Physics, University of Melbourne, Parkville 3010, VIC, Australia}
 \and
 \textit{ARC Centre of Excellence for All Sky Astrophysics in 3 Dimensions (ASTRO 3D), Australia}
\and
\textit{Space Telescope Science Institute, 3700 San Martin Drive, Baltimore MD, 21218
}
\and \textit{Center for Astrophysical Sciences, Department of Physics and Astronomy, Johns Hopkins University, Baltimore, MD, 21218, USA}
\and \textit{Cosmic Dawn Center (DAWN), Denmark}
\and \textit{Niels Bohr Institute, University of Copenhagen, Jagtvej 128, 2200 København N, Denmark}
\and \textit{IPAC, California Institute of Technology, MC 314-6, 1200 E. California Boulevard, Pasadena, CA 91125, USA}
\and \textit{Centre for Astrophysics and Supercomputing, Swinburne University of Technology, PO Box 218, Hawthorn, VIC 3122, Australia}
\and \textit{Minnesota Institute for Astrophysics, University of Minnesota, 116 Church Street SE, Minneapolis, MN 55455, USA}
\and \textit{School of Astronomy and Space Science, University of Chinese Academy of Sciences (UCAS), Beijing 100049, China}
\and \textit{National Astronomical Observatories, Chinese Academy of Sciences, Beijing 100101, China}
\and \textit{Institute for Frontiers in Astronomy and Astrophysics, Beijing Normal University, Beijing 102206, China}
\and \textit{School of Earth and Space Exploration, Arizona State University,
Tempe, AZ 85287-1404, USA}
\and \textit{Department of Mathematics and Physics, University of Ljubljana, Jadranska ulica 19, SI-1000 Ljubljana, Slovenia}
\and \textit{Department of Physics and Astronomy, University of California, Davis, 1 Shields Ave, Davis, CA 95616, USA}
}

 \date{Accepted 20 February 2023 / Received 9 January 2023}
 
\abstract{The escape fraction of Lyman-continuum (LyC) photons ($f_{esc}$) is a key parameter for determining the sources of cosmic reionization at $z\geq 6$. At these redshifts, owing to the opacity of the intergalactic medium, the LyC emission cannot be measured directly. However, LyC leakers during the epoch of reionization could be identified using  indirect indicators
that have been extensively tested at low and intermediate redshifts. 
These include a high \oiii/\oii\ flux ratio, high star-formation surface density, and compact sizes.   
In this work,  we present  observations  of 29 $4.5 \leq z \leq 8$ gravitationally lensed galaxies in the  Abell 2744 cluster field. From a combined analysis of \jwst-\NIRSpec\ and \NIRCam\  data, we accurately derived their
physical and spectroscopic properties: our galaxies  have low masses $(\log(M_\star)\sim 8.5)$, blue UV spectral slopes ($\beta \sim -2.1$), compact sizes ($r_e \sim 0.3-0.5$  kpc), and high \oiii/\oii\ flux ratios. We confirm that these properties are similar to those characterizing low-redshift LyC leakers. Indirectly inferring the fraction of escaping ionizing photons, we find that more than 80\% of our galaxies have predicted $f_{esc}$ values larger than 0.05, indicating that they would
be considered leakers. The average predicted $f_{esc}$ value of our sample is 0.12, suggesting that similar galaxies at $z\geq 6$ have provided a substantial contribution to cosmic reionization. 
}


 \maketitle
%-------------------------------------------------------------------
\section{Introduction} \label{sec:intro}
The Lyman-continuum (LyC, $\lambda<912$ \AA) photons escaping from star-forming galaxies into the neutral intergalactic medium (IGM) can account for the photon budget required to complete reionization only if a substantial fraction of them escape from the galaxies' interstellar and circumgalactic media (ISM and CGM). Given the number density of star-forming galaxies in the Epoch of Reionization (EoR), an average LyC escape fraction ($f_{esc}$) of $\sim 10\%$ across all galaxies would be required \citep[e.g.,][]{Finkelstein2019, Robertson2015} to both reionize the Universe by $z=6$ and match the Thomson optical depth of electron scattering observed in the cosmic microwave background \citep[CMB,][]{Planck2020}. 

However, at $z \geq 4.5$ it is impossible to detect the LyC photons escaping from galaxies, since they are absorbed and scattered by the IGM along the line of sight \citep{inoue2014}. Therefore, efforts at low redshift, where LyC can be detected, have been focused on identifying other observable properties that trace physical conditions facilitating the escape of LyC photons. These  indirect indicators could then be used 
in the EoR to identify the cosmic ionizers. Several indirect diagnostics have been proposed \citep[e.g.,][]{yamanaka2020,Izotov2018b,Marchi_2018,Verhamme2017}, but they are all characterized by a large scatter. One of the best indicators is the presence of a strong Ly$\alpha$ emission \citep[e.g.,][]{Pahl2021, Gazagnes2020}, often characterized by two emission peaks with a small velocity separation. However, at $z>6.5,$ this line is attenuated due to its resonant nature and by the increasingly neutral IGM as we approach the EoR \citep[][]{Pentericci_2018b,mason2019,jung2020,Ouchi_2020,bolan2021}. 
Emission from the \mgii\ doublet has been proposed as a promising LyC proxy \citep{chisholm2020}, as the escape of this line is controlled by resonant scattering in the same low column-density gas as the \lya\ \citep[see also][]{xu2022, izotov2022}.
\\
More recently, the nebular C \textsc{iv} emission line, requiring comparably high ionization energies to the He \textsc{ii} line (E $>$ 47.9\,eV and $>$ 54.4\,eV respectively), has attracted attention since its presence might be strongly linked to the escape of Lyman continuum photons from galaxies \citep[e.g.,][]{Schaerer_2022, Senchyna2022, Saxena2022, Mascia2023}, although this line is in general much fainter than \lya.
Another very popular indicator is the  emission line ratio \oiiidoub/\oiilam\ (hereafter, $O32$), as  \cite{Nakajima2014} first found evidence for its correlation with $f_{esc}$: high values of $O32$ would reflect partially incomplete \hii\ regions, where some LyC photons could escape from \citep[][]{Marchi_2018}. Later on, it was found that this correlation is characterized by a substantial scatter, as highlighted by low-redshift studies \citep[e.g.,][]{Izotov2018b, Nakajima2020, Flury2022}. As a result, a high $O32$ flux ratio is still a necessary condition for a significant measurement of $f_{esc}$, although it is not sufficient by itself to define a LyC leaker, as viewing angles might play a role as well as variation in metallicity and ionization parameter \citep[e.g.,][]{Bassett2019, Katz2020}.
Further additional properties, such as low values of Balmer lines' rest-frame equivalent widths ($EW_0$), are thus required. As a matter of fact, measuring low values of EW in these lines could indicate lower optical depth in the \hii\ region \citep[e.g.,][]{bergvall2013}. However, many local LyC emitters exhibiting high values of Balmer emission line $EW_0s$ are present in the literature \citep{Izotov2016a, Izotov2016b, Izotov2018a, Izotov2018b}. 
A possible explanation for this discrepancy lies in the fact that Balmer-line $EW_0s$ are sensitive not only to \hii\ region size, but also to starburst age and star formation history \citep[][]{Zackrisson2017, Binggeli2018, Alavi2020}. It is thus imperative to pair the EW diagnostic with another indicator in order to prevent degeneracy.
Finally, another diagnostic for a high $f_{esc}$ is the SFR surface density ($\Sigma_{SFR}$): feedback from star formation can blow bubbles or chimneys into the host galaxy's ISM, linking a high $\Sigma_{SFR}$ value to a high $f_{esc}$. This connection is supported by the detection of some compact LyC emitters \citep[LCEs, e.g.,][]{Izotov2018b, Marchi_2018}, even though compactness may not be a defining characteristic of all of them \citep[e.g.,][]{Marchi_2018}. 

Recently, \cite{Flury2022} presented the first statistical test of many of the diagnostics just described on a large sample of low-redshift LCEs. Their conclusion is that indicators based on Ly$\alpha$ emission properties (such as peak velocity separation and equivalent width) perform well, and that diagnostics such as the \oiiialone/\oii\ flux ratio and the SFR surface density, $\Sigma_{SFR}$, could also be used to determine if a galaxy is an LCE or not.

% QUESTO PEZZO NON MI PIACE MOLTO As LyC cannot be directly detected, setting constrains to these indicators remains one of the most controversial topics in modern astronomy. Furthermore, ground-based telescopes are limited in their ability to detect high redshift galaxies in the wavelength range where these indirect indicators might occur. 

With the advent of \jwst, we now have the opportunity to constrain LyC diagnostics during the epoch of reionization with the first \NIRSpec\ observations of high-redshift galaxies. In this paper, we make use of \NIRSpec\ spectra obtained in the Abell 2744 cluster region for a sample of galaxies at $4.5 \leq z \leq 8$ observed as part of the \jwst\ Early Release Science program GLASS \citep[][]{TreuGlass2022} and the \jwst\ Director Discretionary Time program \citep[JWST-GO-2756; P.I. Chen;][]{Roberts-Borsani2022b}. With the wavelength coverage of $0.9 - 5.3\ \mu$m offered by the \NIRSpec\ observations, we can  confirm not only the redshifts of tens of photometrically selected candidates using intense emission lines, but we can also measure optical line ratios and rest-frame EWs with high precision. Our unique data set also includes deep \jwst/\NIRCam\ images, enabling us to better characterize those cosmic reionizer candidates based on their physical properties. 

This paper is organized as follows. We present the data set in Sect.~\ref{sec:data}. We characterize the selected sample and compare the physical and spectroscopic proprieties with cosmological models and other samples at lower redshifts in Sect.~\ref{sec:analysis}. In Sect.~\ref{sec:predictions}, we indirectly infer $f_{esc}$ for the high-redshift sample and in Sect.~\ref{sec:results}, we summarize our key conclusions. Throughout this work, we assume a flat $\Lambda$CDM cosmology with $H_0$ = 67.7 km s$^{-1}$ Mpc$^{-1}$ and $\Omega_m$ = 0.307 \citep{Planck2020} and the \cite{Chabrier2003} initial mass function. All magnitudes are expressed in the AB system \citep{Oke1983}.

\section{Observations and method} \label{sec:data}

The final target list of the GLASS-JWST program and the way in which targets have been prioritized will be presented in a future paper. However, \cite{Morishita2022} have already described the observations and data reduction strategy. Here, we present a brief summary, highlighting the points that are most relevant for this work and describing the methods used to study the properties of the galaxies in our sample. 

\subsection{JWST/NIRSpec MSA observations and data reduction}

\begin{figure*}[ht!]
\includegraphics[width=\textwidth]{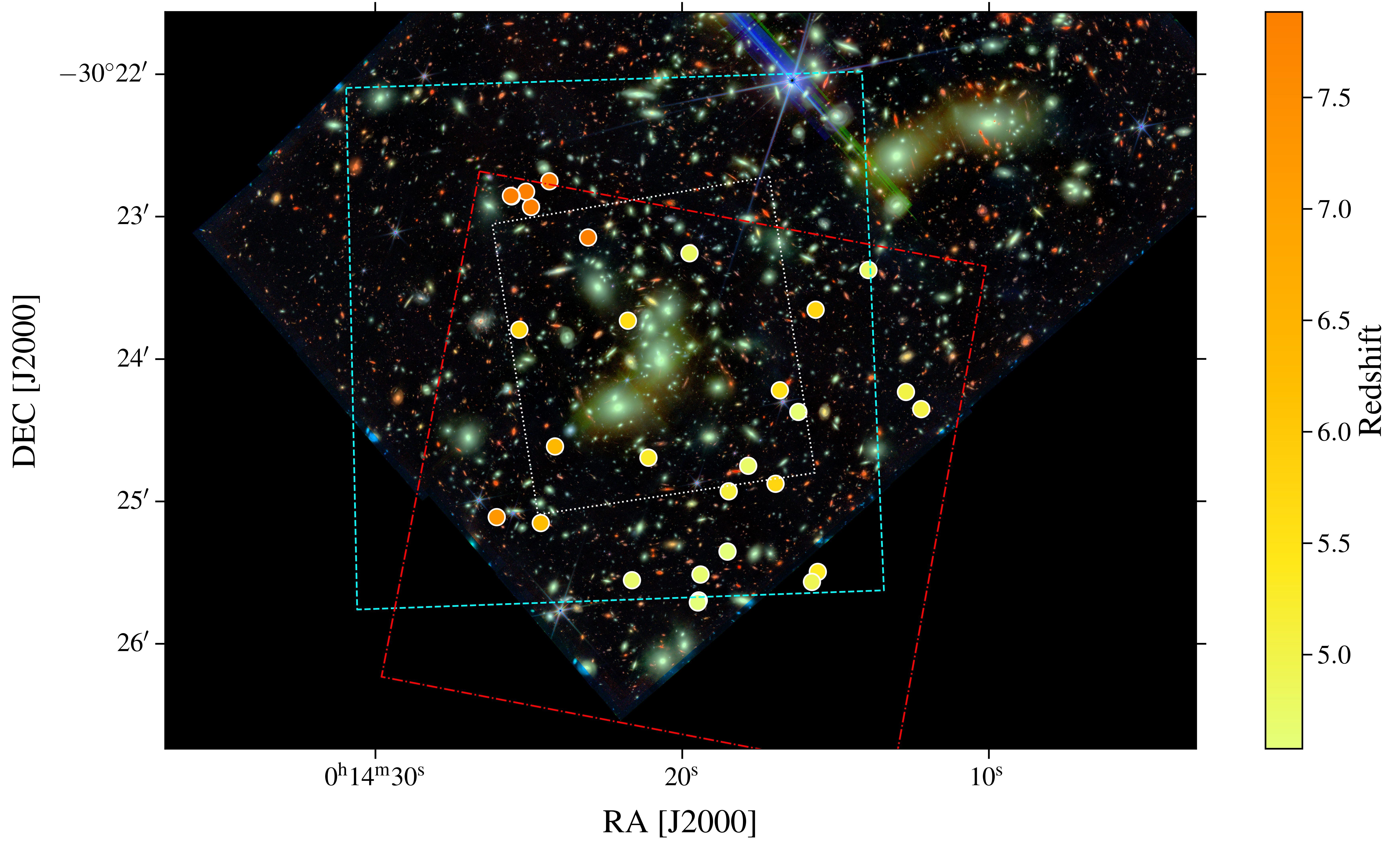}
\caption{Spatial location of the 29 selected sources, color coded by their spectroscopic redshift. They are superimposed on the RGB image of the UNCOVER program, made with NIRCam filters (blue = F115W + F150W, green = F200W + F277W, and red = F356W + F410M + F444W). The MUSE footprint is shown in white, the \NIRSpec\ GLASS-JWST pointing is shown in cyan, and the \NIRSpec\ DDT pointing is shown in red.
\label{fig:footprint}}
\end{figure*}

Our spectra were acquired through \NIRSpec\ MSA observations in two programs: the GLASS-JWST Early Release Science Program \citep[PID 1324, PI Treu;][]{TreuGlass2022} and a JWST DDT program \citep[PID 2756, PI. W. Chen;][]{Roberts-Borsani2022b}, which obtained \NIRSpec\ observations for a subset of targets residing over the central regions of the Frontier Field galaxy cluster Abell 2744. 

The GLASS-JWST observations were carried out on November 10, 2022, with three spectral configurations (G140H/F100LP, G235H/F170LP, and G395H/F290LP). These configurations cover wavelengths between $1-5.14 \ \mu$m, at $R\sim2000-3000$. We exposed each of the three high-resolution gratings for a total of 4.9 hours. Specifically, in this work, we use the G235H/F170LP and G395H/F290LP observations, which contain the bright emission lines we will analyze.

DDT \NIRSpec\ observations were carried out on October 23 2022, using the CLEAR filter+PRISM configuration, which provides continuous wavelength coverage of $0.6-5.3 \ \mu$m at $R \sim 30-300$ spectral resolution. The on-source exposure time is 1.23 hours.

Data were reduced using the official STScI \jwst\ pipeline (ver.1.8.2)\footnote{\url{https://github.com/spacetelescope/jwst}} for Level 1 data products and the \texttt{msaexp}\footnote{\url{https://github.com/gbrammer/msaexp}} code for Level 2 and 3 data products, which is based on the STScI pipeline but also includes additional correction routines. In summary, we initially reduced the uncalibrated data using the \texttt{Detector1Pipeline} routine and the latest set of reference files (jwst\_1023.pmap) to correct for detector-level artifacts and convert them to count-rate images. Then, we applied custom preprocessing routines from \texttt{msaexp} to remove residual $1/f$ noise that is not corrected by the IRS2 readout, to identify and remove "snowballs", and to remove bias exposure by exposure before running STScI routines from \texttt{Spec2Pipeline} for the final 2D cutout images. To perform WCS registration, flat-fielding, path-loss corrections, and flux calibration, these routines include \texttt{AssignWcs}, \texttt{Extract2dStep}, \texttt{FlatFieldStep}, \texttt{PathLossStep}, and \texttt{PhotomStep}. Of note, our chosen reference files include an in-flight flux calibration, accounting for \NIRSpec's better-than-expected throughput at blue wavelengths. Local background subtraction was performed using a three-shutter nod pattern before the resulting images are drizzled onto a common grid. We optimally extracted the spectra using an inverse-variance weighted kernel, which is derived by summing the 2D spectrum along the dispersion axis and fitting the signal along the spatial axis to a Gaussian profile. We visually inspected all kernels to make sure spurious events are not included. As a result, the kernel extracts the 1D spectrum along the dispersion axis. The final step was to verify the default wavelength calibration for the gratings, which is accurate within 1 \AA\ (Williams et al. in prep). 

\subsection{Imaging data}

Deep \NIRCam\ images were acquired from the GO program UNCOVER (GO 2561; PI I. Labbe)\ and included observations in the F115W, F150W, F200W, F277W, F356W, F410M, and F444W filters. The imaging data were reduced using the STScI JWST pipeline and the latest versions of photometric zero points and reference files. A detailed description of all the reduction and calibration steps is presented in \cite{Merlin2022}. Of the 29 sources analyzed in this work (see next section), 27 were observed within the UNCOVER pointings. Their positions and the UNCOVER footprints are presented in Fig. \ref{fig:footprint}. 

\subsection{Emission line and redshift identifications}
The focus of the study is on all sources at $z \geq 4.5$. Specifically, we analyzed the spectra of: 13 galaxies with a spectroscopic redshift larger than 4.5 previously confirmed by the MUSE observations \citep{mahler2018,richard2021}, all showing Ly$\alpha$ in emission in their optical spectra (the footprint of the MUSE observations is also shown in Fig. \ref{fig:footprint}); 29 galaxies with a photometric redshift in the range 4.5-8, of which 5 were selected as part of the $z \simeq 7.9$ candidate protocluster and whose confirmation has been recently presented by \cite{Morishita2022}.
Finally, 23 of the 42 galaxies were observed as part of GLASS-JWST, while 19 were part of the DDT program. 

\begin{figure*}[ht!]
\centering
\includegraphics[width=0.9\textwidth]{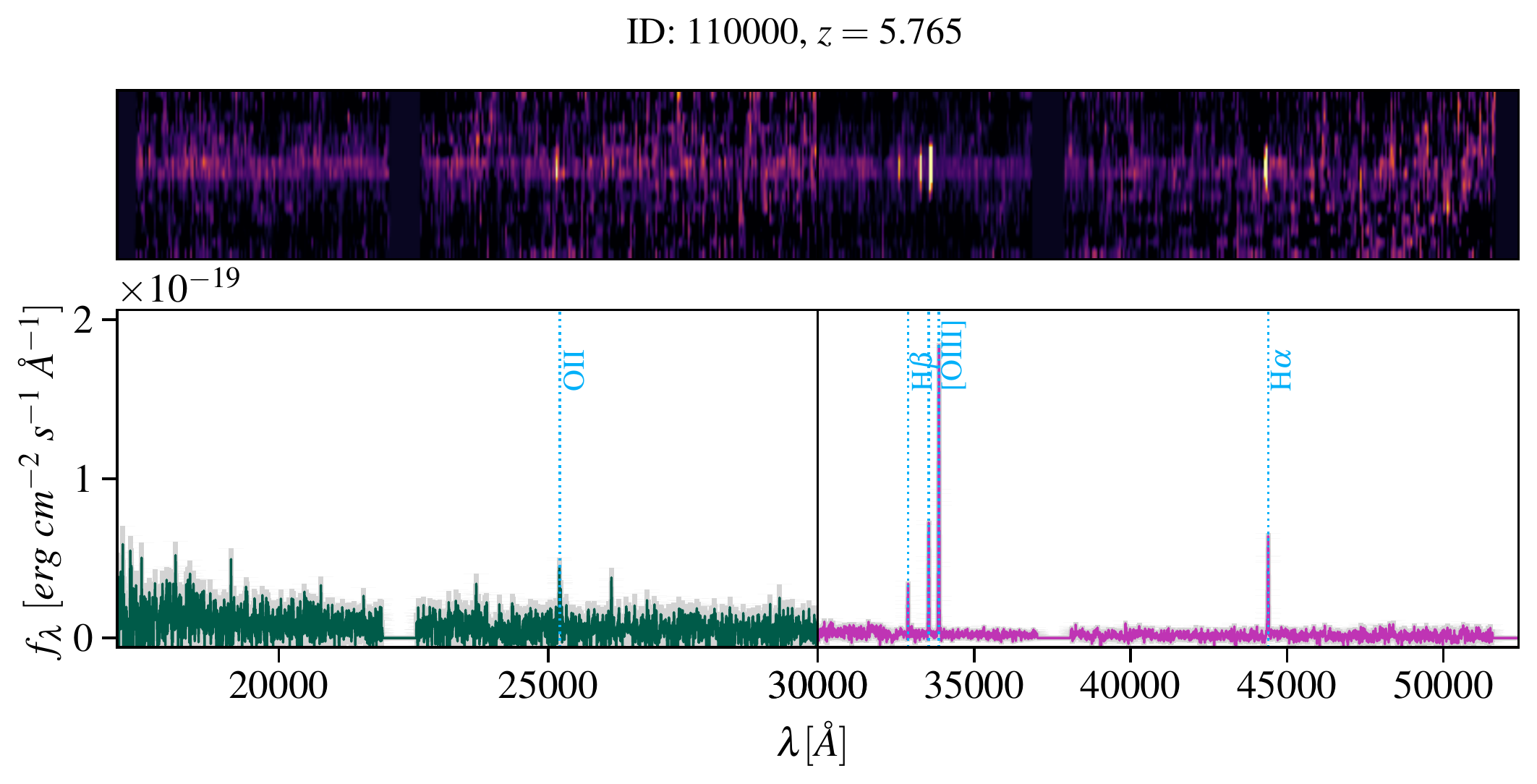}
\includegraphics[width=0.9\textwidth, trim = 0 0 0.8cm 0] {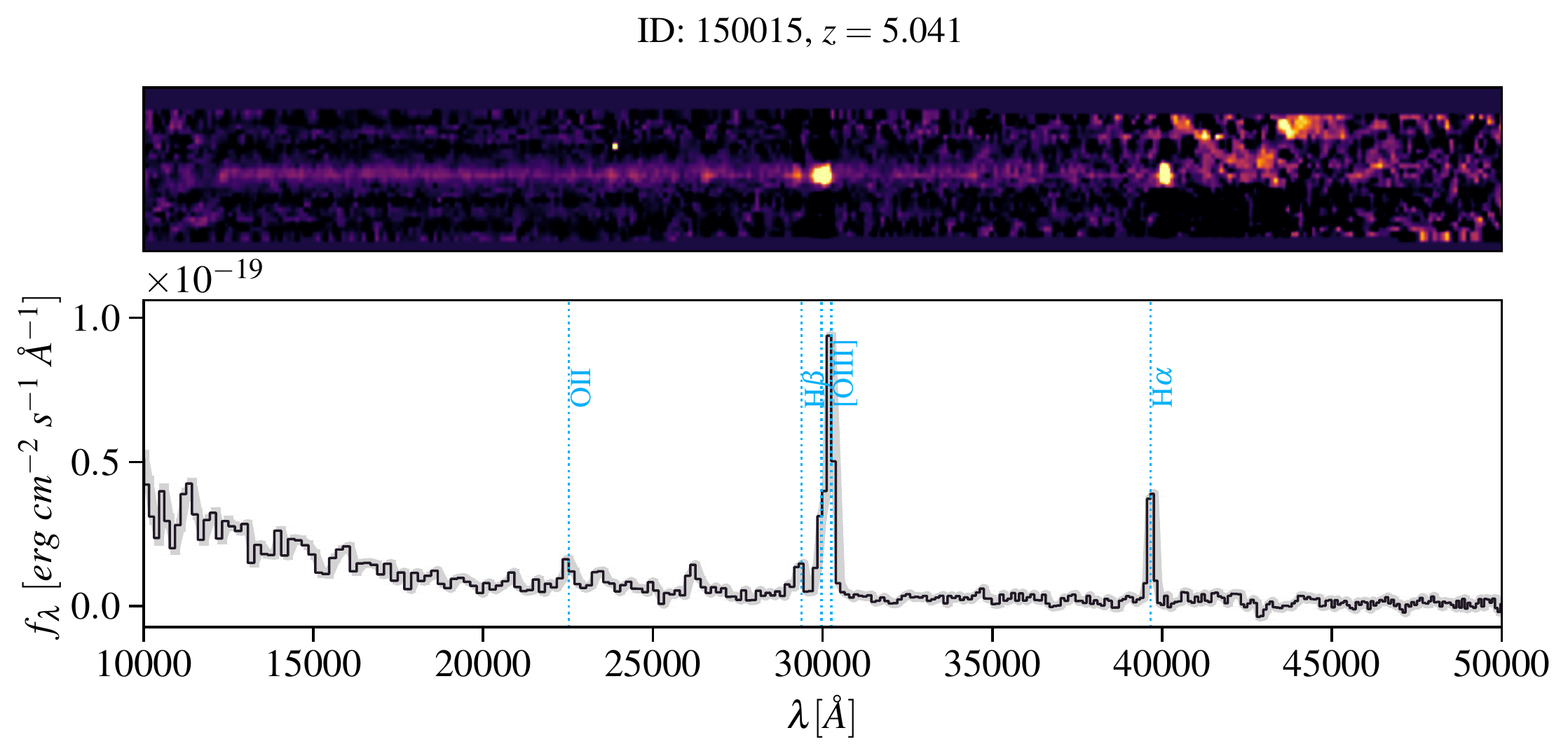}
\caption{Example 2D and 1D spectra of two representative galaxies in our sample. Top: 2D (top) and 1D (bottom) \NIRSpec\ GLASS-JWST spectrum of 110000 at $z = 5.765$ (green for the G235H/F170LP and purple for the G395H/F290LP configuration) and 1$\sigma$ uncertainty (gray). Bottom: 
2D (top) and 1D (bottom) \NIRSpec\ DDT spectrum of 150015 at $z = 5.041$ (black) and 1$\sigma$ uncertainty (gray). Vertical dashed lines mark the position of well-detected rest-optical emission lines. The continuum emission is also detected as seen in the 2D spectrum. On the X-axis in the bottom panel, the observed wavelength (\AA) is reported.
\label{fig:GLASS_50038}}
\end{figure*}

Spectra were visually examined for detectable optical lines using the spectroscopic or photometric redshift information. For photometric sources we determined the spectroscopic redshift when possible using the \hb, \oiiidoub,\ and (when present) \ha\ lines. 
In 29 cases, the \oiilam, \oiiidoub\ and \hb\ were detected and their line fluxes were measured. We also measured the \ha\ line in 17 out of 29 cases, since it falls within the observed spectrum (examples are shown in Fig. \ref{fig:GLASS_50038}). For this part of our analysis, we used the latest version of the \texttt{specutils}\footnote{\url{https://specutils.readthedocs.io/en/stable/index.html}} packages in \textsc{python}. 

From our initial target list of 42 galaxies, there were eight sources with a spectroscopic redshift confirmed from previous MUSE observations, which were placed in the MSA, but for which we are unable to confirm any emission line from JWST data. Of these, five were observed as part of GLASS-JWST and three during the DDT. Most of these sources are extremely faint (fainter than 28 F150W), thus, their redshift confirmation was solely based on the presence of faint \lya\ emission (flux on the order of $1-3 \times 10^{-19} \ \text{erg}/\text{s}/\text{cm}^2$) in the MUSE cubes.
We were also unable to detect any emission lines for five galaxies with a photometric redshift between 4.5 and 7: these galaxies are also faint ($m_{\text{F150W}} \simeq 27.2-28$). However, in these cases it is possible that the photometric redshift was incorrect and this is why we are unable to confirm any emission line in their spectra. 

In Table \ref{tab:summary}, we report the GLASS-JWST IDs, the coordinates and the spectroscopic redshifts of all sources together with their $m_{\text{F150W}}$ magnitudes derived from the imaging data when present.

\subsection{Dust correction and flux measurements} 
We measured the total flux of all detected lines (Balmer lines, \oii,\ and of \oiii) with a Gaussian fit of each line component. From the flux measurement we subtracted a constant continuum emission, which is estimated as the signal averaged over regions in which there are no lines near the one being measured. When the signal-to-noise ratio (S/N) of \oii\ or \hb\  $\leq 2$, we set 2$\sigma$ as an upper limit. 
Prior to carry out a quantitative analysis, it is necessary to consider corrections for dust reddening. For 22 galaxies, \ha\ and \hb\ are both available: for these, we calculated the correction for dust extinction on the basis of the Balmer decrement, assuming a \citet[][]{Calzetti2000} extinction law and an intrinsic ratio $\ha/\hb = 2.86$ \citep[see e.g.,][]{Dominguez2013, Kashino2013, Price2014}, which is valid for an electric temperature of 10000 K. For the other 7 sources where \ha\ is not in the observed range, we applied the average correction $E(B-V)_{neb} \sim 0.1$ derived from all other sources. 

With the dust corrected values, we calculated the $R23 = (\oiiialone + \oii)/\hb$ and $O32$ line fluxes ratios and the \hb\ rest-frame $EW_0$s. We list all these values in in Table \ref{tab:summary}. It is important to highlight that the $O32$ values slightly depend on the dust correction. The assumed temperature may be too low for high-z star-forming galaxies \citep[e.g.,][]{curti2023, nakajima2023}, however the \ha/\hb\ ratio varies by less than 5\% for a 20000 K temperature \citep{Dopita2003}
and therefore the $O32$ variation are well below the current uncertainties.

\begin{table*}
\caption{Sample and spectroscopic measurements.}\label{tab:summary}
$$
\begin{array}{llccccrrr}
\hline \hline
\noalign{\smallskip}
\text{PROGRAM} & \text{GLASS ID} & \text{RA} & \text{DEC} & z_{spec} & m_{\text{F150W}} & EW_0(\hb) & R23 & O32 \\
 & & [\text{deg}] & [\text{deg}] & & [\text{mag}] & [$\AA$] & & \\
\noalign{\smallskip}
 \hline
 \noalign{\smallskip}
\textbf{DDT} &
 10025 & 3.59609 & -30.38581 & 7.875 & 26.7 & 139 \pm 30 & 8 \pm 3 & 6.6 \pm 1.4\\
 &40196 & 3.55803 & -30.38962 & 4.760 & 25.5 & 167 \pm 42 & > 11 & > 90 \\
 & 50017 &3.57006 &-30.40369 &5.550 & 26.3 & 89 \pm 17 & 8 \pm 3 & 9 \pm 2\\
 & 70002 & 3.60544 & -30.3966 & 5.771 & 27.0 & >210 & > 12 & > 8\\
& 80075 & 3.56755 & -30.40623 & 4.632 &26.9 & 173 \pm 38 & 3.4 \pm 1.5 & 9 \pm 2\\
%& 100002 & 3.60338 &-30.38223 & 7.877 & 26.4 & &\\
& 100004 & 3.60657 & -30.38093 & 7.884 & 26.9 & > 130  & > 5 & > 5\\
& 110003 & 3.59069 & -30.39554 & 5.660 & 25.1 & 99 \pm 19 & > 4 & > 8\\
& 150015 & 3.55085 & -30.40590 & 5.041 & 25.7 & 268 \pm 53 & 4.1 \pm 1.8 & 7.2 \pm 1.9\\
& 150053 &3.58120 &-30.42853 &4.580 & 26.7 & 212 \pm 51 & > 5 & > 21\\
& 160170 & 3.56570 & -30.42613 & 4.895 & - & 190 \pm 43 & 5 \pm 2 & 8 \pm 2\\
& 160281 & 3.59015 & -30.42593 & 4.715 & 25.6 & 231 \pm 52 & 5 \pm 2 & 5.3 \pm 1.4\\
& 160284 & 3.58083 & -30.42526 & 4.700 & 24.8 & 130 \pm 27 & > 5 & > 8\\
& 160345 & 3.55293 & -30.40389 & 5.020 & 27.2 & 110 \pm 44 & > 4 & > 5\\
\noalign{\smallskip}
\hline
\noalign{\smallskip}
\textbf{GLASS-JWST} &
10000 & 3.60134 & -30.37923 & 7.884 & 25.5 & 76 \pm 13 & 7 \pm 2 & 8 \pm 3\\
& 10021 & 3.60851 & -30.41854 & 7.288 & 25.4 & 104 \pm 24 & 12 \pm 5 & 13 \pm 5\\
& 50002 & 3.57700 & -30.41552 & 5.135 & 27.8 & 69 \pm 23 & 12 \pm 6 & 19 \pm 7 \\
& 50038 & 3.56520 & -30.39426 &5.773 & 25.5 & 92 \pm 19 & 10 \pm 4 & 9 \pm 3\\
%contaminated by a low z source 70003 & 3.57585 & -30.38929 & 5.622 && 53.7 \pm 27.9 & \textbf{NO LINES}\\
& 70018 & 3.58790 & -30.41159 & 5.284 & 27.1 & 155 \pm 32 & > 4 & > 12\\
& 80070 & 3.58232 & -30.38765 & 4.798 & 26.4 & 135 \pm 25 & 12 \pm 5 & 53 \pm 20\\
& 80085 & 3.57435 &-30.41253 & 4.728 & 28.1 ^a & 168 \pm 24 & 2.5 \pm 1.5 & 19 \pm 11\\
& 100001 & 3.60385 & -30.38223 & 7.875 & 26.5 & 39 \pm 8 & 10 \pm 4 & 3.2 \pm 1.0\\
& 100003 & 3.60451 & -30.38044 & 7.880& 26.2 & 85 \pm 18 &11 \pm 4 & 21 \pm 7\\
& 100005 & 3.60646 & -30.38099 & 7.883& 26.7 & 33 \pm 15 & 21 \pm 12 & 2.9 \pm 1.0\\
& 110000 & 3.57064 & -30.41464 & 5.765 & 25.8 & 57 \pm 13 & 9 \pm 4 & 6 \pm 2\\
& 150008 &3.60253& -30.41923& 6.230 & 25.8 & 141 \pm 30 & > 7 & > 20\\
& 150029 & 3.57717 & -30.42258 & 4.585 & 26.1 & 205 \pm 45 & 8 \pm 3 & 17 \pm 6\\
& 160122 & 3.5649 & -30.42496 & 5.333 & - & 149 \pm 25 & 5 \pm 2 & 18 \pm 7\\
& 160275 &3.58107 &-30.42830 &4.578 & 26.2 & 5 \pm 1 & - & - \\
& 400009 &3.60059 &-30.41027& 6.376 & 26.8 & 35 \pm 7 & - & -\\
\noalign{\smallskip}
\hline
\end{array}
$$
\begin{tablenotes}
 \small
 \item $^a$ from MUSE catalog, magnitude F140W HST. \\ Missing O32 and R23 values are due to \oii\ undetection because of its position in un-acquired part of the spectrum. \hb\ and \oiii\ are always detected.
 \end{tablenotes}
\end{table*}

\subsection{Measurements of physical parameters }\label{sec:SED_fitting}

Following \cite{Santini2022}, we measured the stellar masses $M_{\star, obs}$, the observed absolute UV magnitudes at 1500\AA\ ($M_{1500, obs}$), the star formation rates (SFRs), and dust reddening $E(B-V)$ by fitting synthetic stellar templates to the seven-band \NIRCam\ photometry and the released HST photometry \citep{Castellano2016} with \textsc{zphot} \citep{fontana00}. We adopted \cite{Bruzual2003} models and assumed delayed exponentially declining star formation histories -- SFH($t$)$\propto (t^2/\tau) \cdot \exp(-t/\tau)$ -- with $\tau$ ranging from 0.1 to 7 Gyr. The age ranges from 10 Myr to the age of the Universe at each galaxy redshift, while metallicity can assume values of 0.02, 0.2 or 1 times Solar metallicity. For the dust extinction, we used the \cite{Calzetti2000} law with $E(B-V)$ ranging from 0 to 1.1. We computed $1\sigma$ uncertainties on the physical parameters by retaining, for each object, the minimum and maximum fitted masses among all the solutions with a probability $P(\chi^2)>32\%$ of being correct, fixing the redshift to the best-fit value. 

As a final step, we needed to adjust the stellar masses and $M_{1500}$ and the SFR values for the lensing amplification factor.
The magnification factor $\mu$ was derived by combining the LM-model \citep{Bergamini2022}, the model used by \cite{Roberts-Borsani2022c}, with a new spatially more extensive model that fully covers the JWST field of view (Bergamini et al. in prep). The $\mu$ values range from 1.6 to 12, with a median value of $\sim 2$.
The results on our sources' stellar masses, $M_{\star, true}$, their $M_{1500}$, and their lensing magnification factors $\mu$ are reported in Table \ref{tab:summary2}.

\subsection{Sizes and SFR surface density estimates}
 
The SFRs were derived from the \ha\ emission line using the standard conversion by \cite{Kennicutt1998}, that is, $SFR(\ha) [M_\odot \ \text{yr}^{-1}]= 7.9 \ 10^{42} \ L_{\ha}$, and then corrected for the Chabrier IMF. For those few sources at $z \geq 6.6,$ where the \ha\ line falls outside the observable window we used  dust-corrected \hb\ fluxes instead. To correct for slit losses and possible residual uncertainties on flux calibration, we normalized the spectra to the F444W filter by integrating the spectrum under the F444W bandpass, the closest to the \ha\ line in our sample. 
Thanks to the high resolution of the GLASS-JWST spectra, no correction for \nii\ contamination is required for \ha\ measurements. For galaxies observed by the DDT programs, \ha\ is blended with the \nii\ doublet. However our sources are all very-low-mass galaxies and, based on low-redshift results, we expect contamination to be less than 10\% \citep[e.g.,][]{Faisst2018}. Therefore, we did not attempt to make any correction on the fluxes.

The results obtained on the SFR were also compared to the values determined by the SED fitting procedure described in Sect. \ref{sec:SED_fitting}, finding a reasonable agreement between the two data sets, indicating that there are no systematic issues. Obviously, the SED fitting SFR is heavily dependent on the choice of SFH, and also on the dust correction, while the value derived from the \ha\ luminosity is sensitive to short timescales. That is to say that it only probes the presence of short-lived, massive stars, so their  ratio is actually an indication of the burstiness of the galaxy.

We measured the size of each galaxy $r_{e}$ in the F115W band, corresponding to the UV rest-frame of the galaxies.
Adopting forward-modeling technique, we assumed that galaxies are well represented by a Sersic profile \citep{sersicpaper} as \cite{Yang2022}. Then we modeled the appearance of the source in the image plane considering lensing and PSF effects.
The source reconstruction was performed via \texttt{python} software 
\texttt{Lenstruction} \cite[see details in][]{Yang_2020}, which is built on \texttt{Lenstronomy} \citep{Birrer2021}.
% , assuming the the galaxies are well represented by a Sersic profile and following \cite{Yang2022}, who used the \texttt{python} software \texttt{Galight}\footnote{\url{https://github.com/dartoon/galight}} \citep{Ding2020}. 
% Source reconstruction from the lensed galaxies considering lensing and psf effect was then done using the tool "Lenstruction" as implemented in the publicly available multi-purpose gravitational lensing software Lenstronomy. Details can be found in \cite{Yang_2020}. 
In this way we obtained the intrinsic properties of galaxies in the source plane, hence, the intrinsic size. Finally, we calculated the SFR surface density using the relation $\Sigma_{SFR} = SFR/2\pi r_{e}^2$ \citep[e.g.,][]{Naidu2020, Flury2022}.
The SFR surface densities $\Sigma_{SFR}$ are listed in Table \ref{tab:summary2}.

\subsection{UV-$\beta$ slopes}
We measured the UV slope of our galaxies from the \NIRCam\ photometry and/or the previously available \hst\ photometry \citep{Castellano2016}, with the approach detailed in \cite{Calabro2021}.
In brief, we considered all the photometric bands whose entire bandwidths are  between a $1216$ and $3000$ \AA\ rest frame. The former limit is set to exclude the \lya\ line and Ly-break, while the latter limit is slightly larger than that adopted in \cite{Calabro2021} to ensure a larger range. 

Then, we fitted the selected photometry with a single power-law of the form f($\lambda$) $\propto \lambda^\beta$. 
In practice, we fitted two or three photometric bands amongst \hst\ F814W and \jwst-\NIRCam\ F115W, F150W or F200W depending on the exact redshift of the sources. This choice allows us to  uniformly probe the spectral range between $1500$ and $3000$ \AA\ for most of the galaxies. 
We measured the $\beta$ and associated uncertainty for each source using a bootstrap method. By using $n = 500$ Monte Carlo simulations, the fluxes in each band were varied according to their error. The results provided a resultant slope distribution from which we calculated the mean and standard deviation of $\beta$ for each galaxy.
The results on $\beta$ with associated errors are reported in Table \ref{tab:summary2}. We find a median $\beta$ of $-2.1$, with a $1\sigma$ dispersion of $0.4$. The UV magnitudes $M_{1600}$ that can be derived simultaneously with this approach are consistent with the values obtained from the SED fitting and described in the previous section.

\begin{table*}
\caption{Physical parameters derived from multi-band photometry. All measurements are corrected for magnification where needed.}\label{tab:summary2}
$$
\begin{array}{llcccccc}
\hline \hline
\noalign{\smallskip}
\text{PROGRAM} & \text{GLASS ID} & \mu^* & \log_{10}(M_\star) & r_e & \log_{10}(\Sigma_{SFR}) & M_{1500} & \beta \\
 & & & [\text{M}_\odot] & [\text{kpc}]& [\text{M}_\odot \ \text{yr}^{-1} \ \text{kpc}^{-2}] & [\text{mag}]\\
\noalign{\smallskip}
 \hline
 \noalign{\smallskip}
\textbf{DDT} &
 10025 & 2.60^{+0.08}_{-0.07} & 8.03 _{- 0.14 }^{+ 0.19 } & 0.40 & 0.4 \pm 0.2 & -19.00 _{- 0.03 }^{+ 0.01 } & -2.08 \pm 0.45 \\
 &40196 & 2.95^{+0.21}_{-0.72} & 7.87 _{- 0.13 }^{+ 0.13 } & 0.33 & 1.6 \pm 1.0 & -19.41 _{- 0.03 }^{+ 0.02 } & -2.03 \pm 0.48\\
 & 50017 & 2.27^{+0.07}_{-0.16} & 8.78 _{- 0.69 }^{+ 0.23 } & 0.43& 0.7 \pm 0.1 & -19.13 _{- 0.19 }^{+ 0.10 } & -1.77 \pm 0.25\\
 & 70002 & 2.19^{+0.05}_{-0.05} & 7.13 _{- 0.15 }^{+ 0.01 } & 0.12 & 2.0 \pm 1.3 & -18.44 _{- 0.01 }^{+ 0.01 } & -2.54 \pm 0.26\\
& 80075 & 2.02^{+0.06}_{-0.17} & 6.76 _{- 0.15 }^{+ 0.23 } & 0.07 & 2.3 \pm 1.7 & -17.37 _{- 0.03 }^{+ 0.01 } & -2.26 \pm 0.23\\
%& 100002 & 3.60338 &-30.38223 & 7.877 & 26.4 & &\\
& 100004 & 1.92^{+0.06}_{-0.08} & 9.17 _{- 0.28 }^{+ 0.14 } & 0.40 & 0.2 \pm 0.3 & -19.76 _{- 0.10 }^{+ 0.11 } & -1.88 \pm 0.44 \\
& 110003 & 11.6^{+0.9}_{-1.0} & 7.92 _{- 0.06 }^{+ 0.71 } & 0.48 & 1.2 \pm 0.5 & -20.52 _{- 0.03 }^{+ 0.15 } & -2.51 \pm 0.24 \\
& 150015 & 1.79^{+0.05}_{-0.29} & 7.87 _{- 0.44 }^{+ 0.34 } & 0.38 & 1.2 \pm 0.7 & -19.52 _{- 0.15 }^{+ 0.09 } & -2.54 \pm 0.49 \\
& 150053 & 1.75^{+0.03}_{-0.14} & 7.85 _{- 0.45 }^{+ 0.48 } & 0.80 & -0.1 \pm 0.6 & -19.26 _{- 0.15 }^{+ 0.08 } & -1.25 \pm 0.21 \\
& 160170 & 1.58^{+0.04}_{-0.13} & - & - & - & - & -\\
& 160281 & 1.96^{+0.03}_{-0.15} & 7.93 _{- 0.08 }^{+ 0.24 } & 1.65 & -0.3 \pm 0.9 & -20.00 _{- 0.11 }^{+ 0.01 } & -2.26 \pm 0.51 \\
& 160284 & 1.85^{+0.03}_{-0.15} & 8.85 _{- 0.10 }^{+ 0.01 } & 0.53 & 1.2 \pm 0.6 & -20.16 _{- 0.01 }^{+ 0.02 } & -2.05 \pm 0.48 \\
& 160345 & 1.86^{+0.05}_{-0.30} & 7.95 _{- 0.81 }^{+ 0.09 } & 0.45& 0.5 \pm 0.1 & -18.67 _{- 0.12 }^{+ 0.06 } & -2.49 \pm 0.49 \\
\noalign{\smallskip}
\hline
\noalign{\smallskip}
\textbf{GLASS-JWST} &
10000 & 2.13^{+0.07}_{-0.11} & 8.11 _{- 0.01 }^{+ 0.27 } & 0.20 & 2.0 \pm 1.3 & -19.71 _{- 0.02 }^{+ 0.10 } & -2.27 \pm 0.46 \\
& 10021 & 2.17^{+0.08}_{-0.03} & 8.70 _{- 0.35 }^{+ 0.11 } & 0.68 &0.8 \pm 0.2 & -21.18 _{- 0.02 }^{+ 0.08 } & -2.25 \pm 0.48 \\
& 50002 & 2.15^{+0.06}_{-0.13} & 8.57 _{- 0.11 }^{+ 0.22 } & 0.70 & -1.4 \pm 2.0 & -18.78 _{- 0.11 }^{+ 0.06 } & -1.66 \pm 0.66 \\
& 50038 & 2.66^{+0.12}_{-0.32} & 9.23 _{- 0.39 }^{+ 0.03 } & 0.48 & -1.7 \pm 2.0 & -19.76 _{- 0.09 }^{+ 0.03 } & -1.79 \pm 0.25 \\
%contaminated by a low z source 70003 & 3.57585 & -30.38929 & 5.622 && 53.7 \pm 27.9 & \textbf{NO LINES}\\
& 70018 & 6.89^{+0.21}_{-0.30} & 8.05 _{- 0.14 }^{+ 0.01 } & 0.19 & 1.9 \pm 1.3 & -18.00 _{- 0.04 }^{+ 0.04 } & -2.02 \pm 0.49 \\
& 80070 & 5.45^{+0.26}_{-0.27} & 7.42 _{- 0.06 }^{+ 0.21 } & 0.59 & 0.4 \pm 0.2 & -18.24 _{- 0.03 }^{+ 0.01 } & -1.92 \pm 0.49 \\
& 80085 & 2.11^{+0.05}_{-0.11} & - & - & - & - & -2.43^a \\
& 100001 & 1.99^{+0.06}_{-0.09} & 9.48 _{- 0.13 }^{+ 0.04 } & 0.50 & 0.34 \pm 0.3 & -20.45 _{- 0.02 }^{+ 0.04 } & -1.63 \pm 0.48 \\
& 100003 & 1.96^{+0.06}_{-0.09} & 8.29 _{- 0.15 }^{+ 0.31 } & 0.15 & 2.3 \pm 1.7 & -20.15 _{- 0.02 }^{+ 0.15 } & -2.51 \pm 0.48 \\
& 100005 & 1.92^{+0.06}_{-0.08} & 8.63 _{- 0.27 }^{+ 0.17 } & 0.25 & 1.5 \pm 1.2 & -19.85 _{- 0.07 }^{+ 0.11 } & -2.55 \pm 0.48 \\
& 110000 & 1.90^{+0.05}_{-0.14} & 9.10 _{- 0.62 }^{+ 0.05 } &0.54 & 0.5 \pm 0.2& -20.13 _{- 0.17 }^{+ 0.04 } & -1.70 \pm 0.23 \\
& 150008 & 2.60^{+0.03}_{-0.04} & 8.41 _{- 0.19 }^{+ 0.35 } & 0.40 & 1.5 \pm 0.7 & -19.33 _{- 0.06 }^{+ 0.14 } & -2.10 \pm 0.25 \\
& 150029 & 1.86^{+0.04}_{-0.14} & 8.55 _{- 0.01 }^{+ 0.07 } & 0.27 & 1.7 \pm 1.0 & -18.75 _{- 0.04 }^{+ 0.01 } & -1.85 \pm 0.20 \\
& 160122 & 1.59^{+0.04}_{-0.13} & - & - & - & - & -\\
& 160275 & 1.75^{+0.03}_{-0.14} & 7.73 _{- 0.14 }^{+ 0.11 } & 0.47 & -1.3 \pm 1.8 & -19.20 _{- 0.02 }^{+ 0.01 } & -2.55 \pm 0.20\\
& 400009 & 9.3^{+1.6}_{-0.6} & 6.77 _{- 0.26 }^{+ 0.54 } & 0.11 & 1.0 \pm 0.3 & -17.05 _{- 0.04 }^{+ 0.14 } & -2.17 \pm 0.25\\
\noalign{\smallskip}
\hline
\end{array}
$$ 
\begin{tablenotes}
 \small
 \item * Median magnification of the lens model by Bergamini (in prep.), calculated at the position of the source. Measurements are associated with 1 $\sigma$ uncertainties. $^a$ measured from HST photometry
\end{tablenotes}
\end{table*}

\section{Results}\label{sec:analysis}

\subsection{Possible AGN contamination}
Our sample was selected solely based on known spectroscopic redshift (via faint and narrow \lya\ emission) or through photometric redshifts. It could therefore be affected by AGN contamination. Since we are interested in searching for candidate LyC emitters amongst the star-forming population, we first consider whether the primary source of ionization might not be star formation, but  AGN activity instead.

\begin{figure}[ht!]
\includegraphics[width=\linewidth]{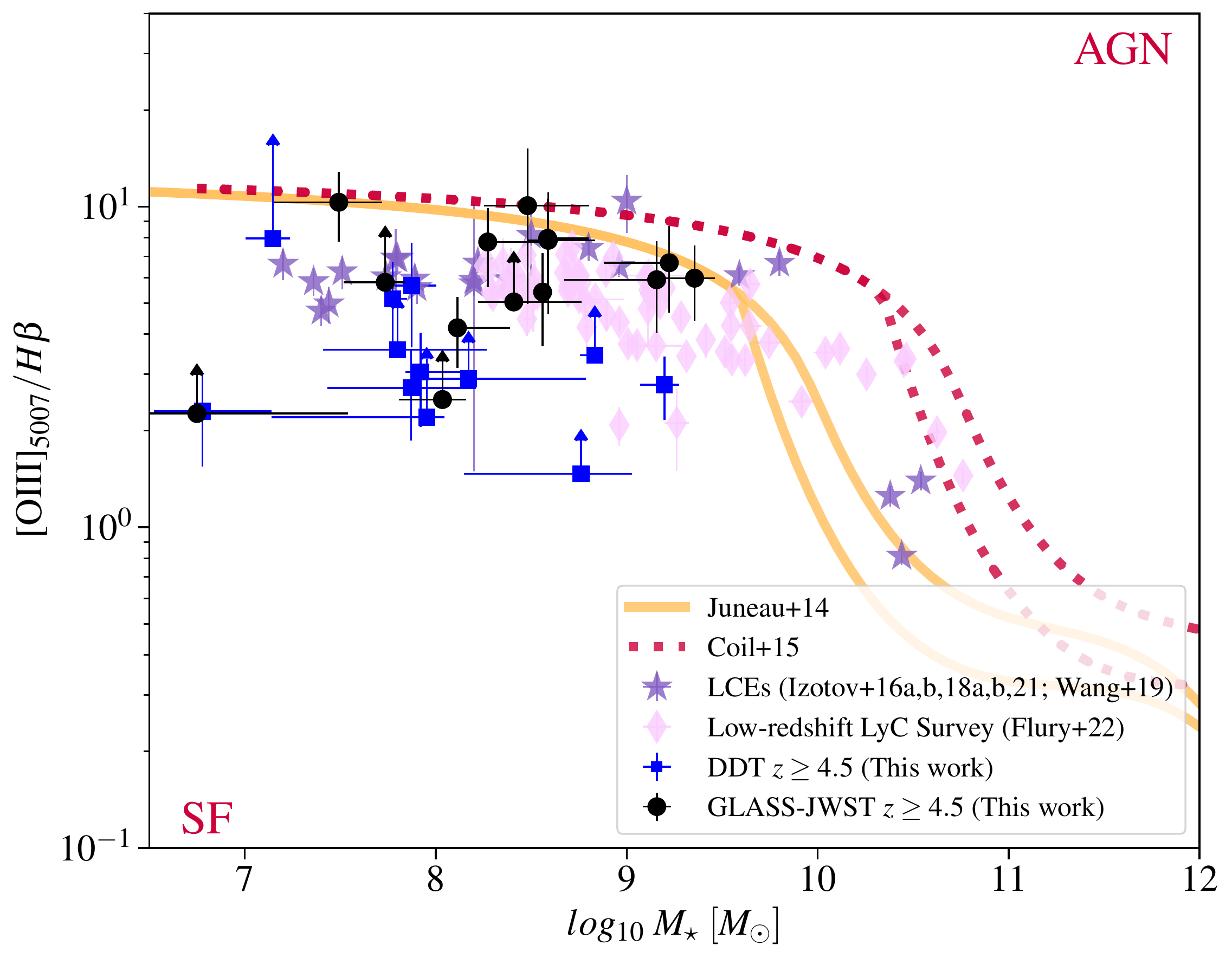}
\caption{MEx diagram for the sample of galaxies analyzed in this paper:black dots and green squares are for the GLASS-JWST and DDT samples, respectively. For reference, we plot also the galaxies at $z = 0.3 - 0.4$ from \cite{Flury2022} (diamonds) and the LCE candidates from previous studies \citep[stars,][]{Izotov2016a, Izotov2016b, Izotov2018a, Izotov2018b, Izotov2021, Wang2019}. The two orange demarcation lines from \cite{Juneau2014} show the boundaries of the AGN-and-star-forming transition region. All objects above the upper line are AGN-dominated; all galaxies below or rightward of the lower line are presumptively dominated by star formation. We also show the separation from \cite{Coil2015} (dotted lines), which is the adaptation of the Juneau et al. model for galaxies and AGNs at $z\sim2.3$ from the MOSDEF survey.
\label{fig:AGN}}
\end{figure}

\begin{figure}[ht!]
\includegraphics[width=\linewidth]{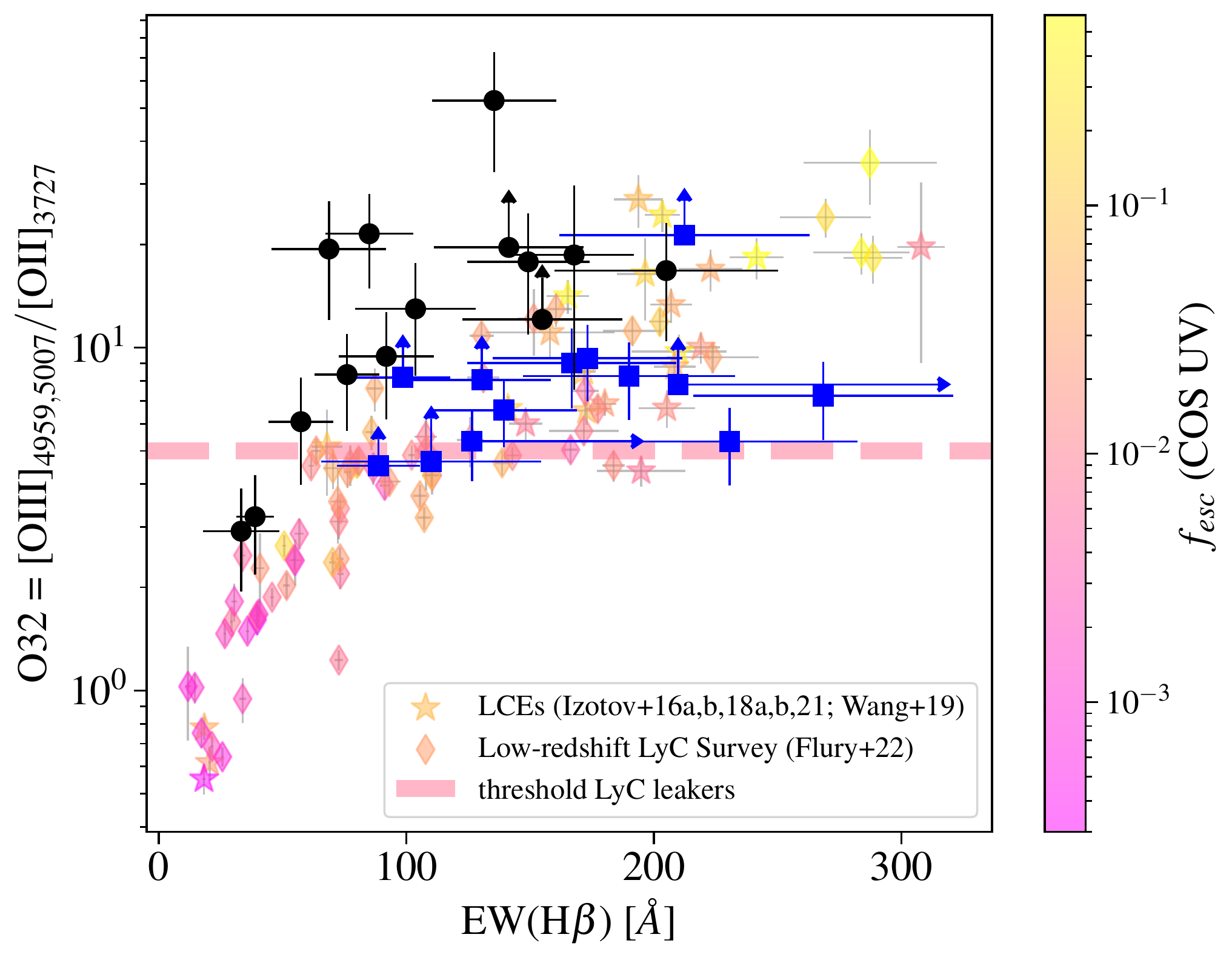}
\caption{$O32$ vs. rest-frame $EW_0(\hb)$. Symbols are the same as in Figure \ref{fig:AGN}. For reference, we also plot  the galaxies at $z = 0.3 - 0.4$ from \cite{Flury2022} (diamonds) and the LCE from \cite{Izotov2016a, Izotov2016b, Izotov2018a, Izotov2018b, Izotov2021, Wang2019} (stars). Symbols are color coded as a function of their measured $f_{esc}$. The pink line ($O32 = 5$) indicates the threshold for LCEs as predicted by \cite{Flury2022}.
\label{fig:O32_Hbeta}}
\end{figure}

 We employed the mass-excitation (MEx) diagram, first proposed by \cite{Juneau2011} to  combine the measurements of the \oiiialone $\lambda 5007$/\hb\ emission line ratio with the stellar mass to discriminate between AGNs and star-forming galaxies. This diagram was proposed as an alternative to the 
 classical BPT diagram that compares the \oiiialone $\lambda 5007$/\hb\ to the \nii/\ha\ emission line ratios \citep[][]{BPT_1981}
 when these last two lines fall out of the visibility window and it is not possible to use them to characterize the ionization mechanism in the galaxies -- as is the case for our low-resolution DDT spectra where the \nii\ doublet is blended with \ha.
Due to the high resolution of the GLASS-JWST spectra, for the galaxies with these observations, we used the dust-corrected flux measurement of the 5007 \AA\ component of the \oiiialone\ doublet. Instead, for the galaxies observed as part of the DDT program for which the doublet cannot be resolved,
we determined the \oiiialone $\lambda 5007$ flux, assuming the expected line ratio of 1:3 for the two components fixed by atomic physics. As shown in Fig. \ref{fig:AGN}, the position of our sources in the MEx diagram indicates that our sample contains essentially star forming galaxies, lying below or around the division line identified by \cite{Coil2015} for $z=2.3$ galaxies and AGN from the MOSDEF survey. For reference, we also plot  the galaxies at $z = 0.3 - 0.4$ from the Low-redshift Lyman Continuum Survey from \cite{Flury2022} and the compilation of low-z LCE also used by the same authors \citep[from][]{Izotov2016a, Izotov2016b, Izotov2018a, Izotov2018b, Izotov2021, Wang2019}.

\subsection{LyC indirect diagnostics}
As discussed above, \lya\ is possibly the best indirect diagnostic of LyC escape, since the conditions that favour the escape of \lya\ photons are often the same that allow for the escape of LyC photons. However, we cannot use this parameter for our sample, since the \NIRSpec\ data do not cover the 1216\AA\ region for most of our galaxies and only a small subset of our sources have been covered by previous MUSE observations (see also Fig. \ref{fig:footprint}). In addition, for the sources at $z \geq 7$, the IGM becomes significantly neutral, thus absorbing the emission even in sources where the line would be intrinsically bright \citep[e.g.,][]{Stark2010,Pentericci2011,Mason2018a}. Indeed,  \cite{Morishita2022} recently showed that none of the galaxies in the protocluster candidate at $z=7.89$ presents bright \lya\ emission and estimate an average neutral hydrogen fraction of the IGM in the region to be $> 0.45$. Therefore in this work we concentrate on the other most promising diagnostics tested at low redshift by \cite{Flury2022} and available for our galaxies: these authors showed that $O32$, $\beta_{1200}$, $r_{50}$, and $\Sigma_{SFR}$ as well as $EW_0(\hb)$ and $M_{1500}$ exhibit some of the strongest and most significant correlations with $f_{esc}$. This would indicate that certain characteristics, such as concentrated star formation, young stellar populations, and high ionization states, play an essential role in the escape of LyC photons.

\begin{figure}[ht!]
\includegraphics[width=\linewidth]{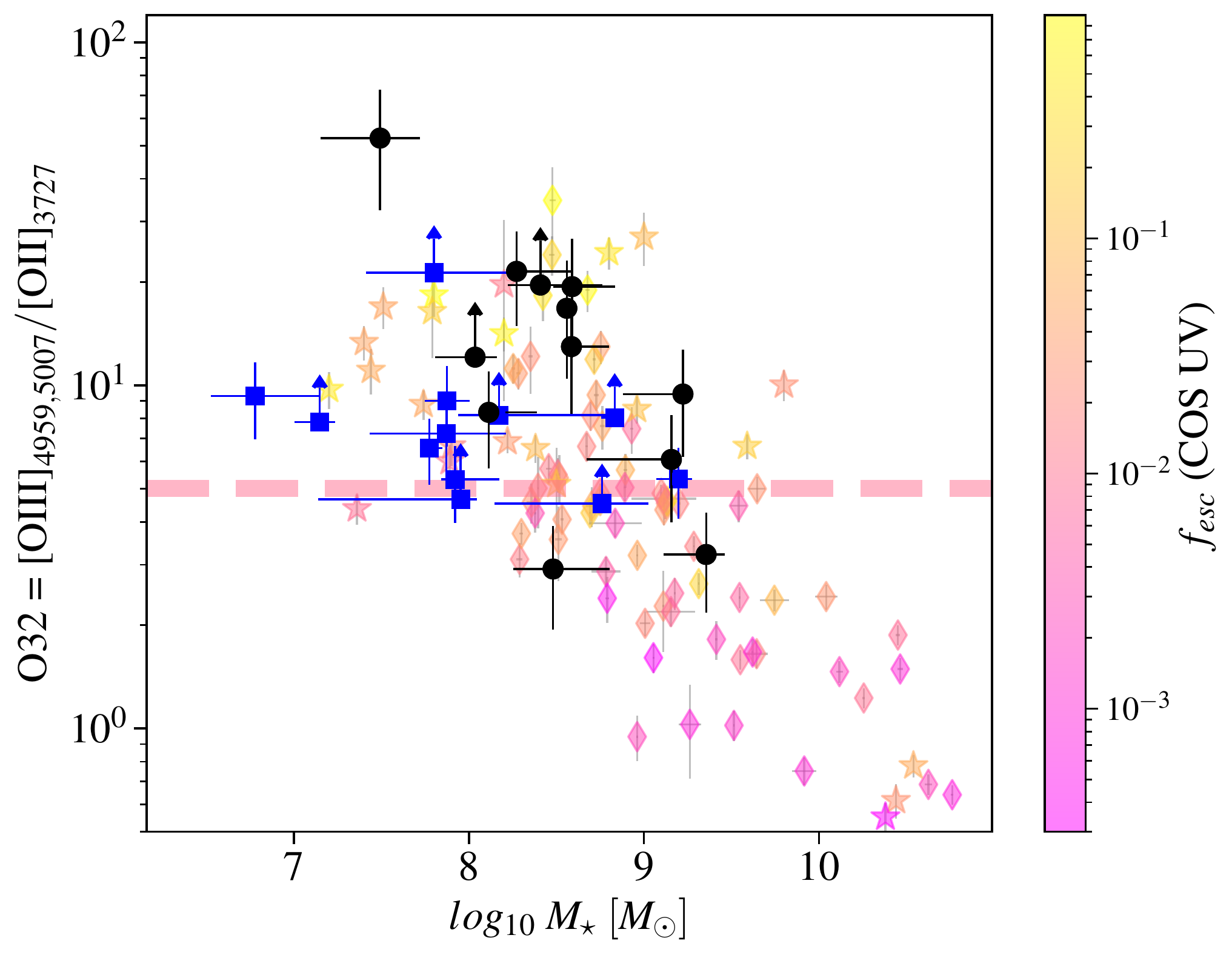}
\caption{$O32$ vs. $M_\star$. Symbols are as in Fig. \ref{fig:O32_Hbeta}.
\label{fig:O32_mass}}
\end{figure}

We began by analyzing the $O32$ and the rest-frame $EW_0(\hb)$ relation, as shown in Fig. \ref{fig:O32_Hbeta}. We plot the values for the \jwst\ high-redshift sample and compare them to the local galaxies with measured $f_{esc}$ from previous works \citep{Flury2022,Izotov2016a, Izotov2016b, Izotov2018a, Izotov2018b, Izotov2021, Malkan2021, Wang2019}. These values are color-coded as a function of their $f_{esc}$ (COS UV) measured values. We can see that the large majority of the high-redshift sources have values of $O32$ larger than 5, which has been indicated as a threshold for LyC leakers (LCEs from here onwards) by \cite{Flury2022}. We note that \cite{Flury2022} define leakers as galaxies with an $f_{esc} > 0.05$ measured with $S/N \geq 5$. Our sample indeed mostly lies in the region of the plot populated by low-redshift leakers. In Fig. \ref{fig:O32_mass}, we show  $O32$ as a function of total stellar mass, again comparing our sample to the low-z galaxies: our sample perfectly overlaps with the low-z galaxies at the low-mass end, where we find most of the LCEs.

\cite{Zackrisson2013} suggested that galaxies ought to be identified with high $f_{esc}$ by combining the UV slope $\beta$ with the measurement of the EW of a Balmer line, such as \hb, and that the predictions are almost independent of the model assumed (i.e., radiation or density bounded nebula). 
We know that both $\beta$ and $EW_0(\hb)$ are also good indirect indicators according to \cite{Flury2022}, although there does not seem to be a direct correlation between the two values: \cite{Flury2022} showed that their LCEs do not seem to follow the predictions provided by the models from \cite{Zackrisson2013}. 

\begin{figure}[ht!]
\includegraphics[width=\linewidth]{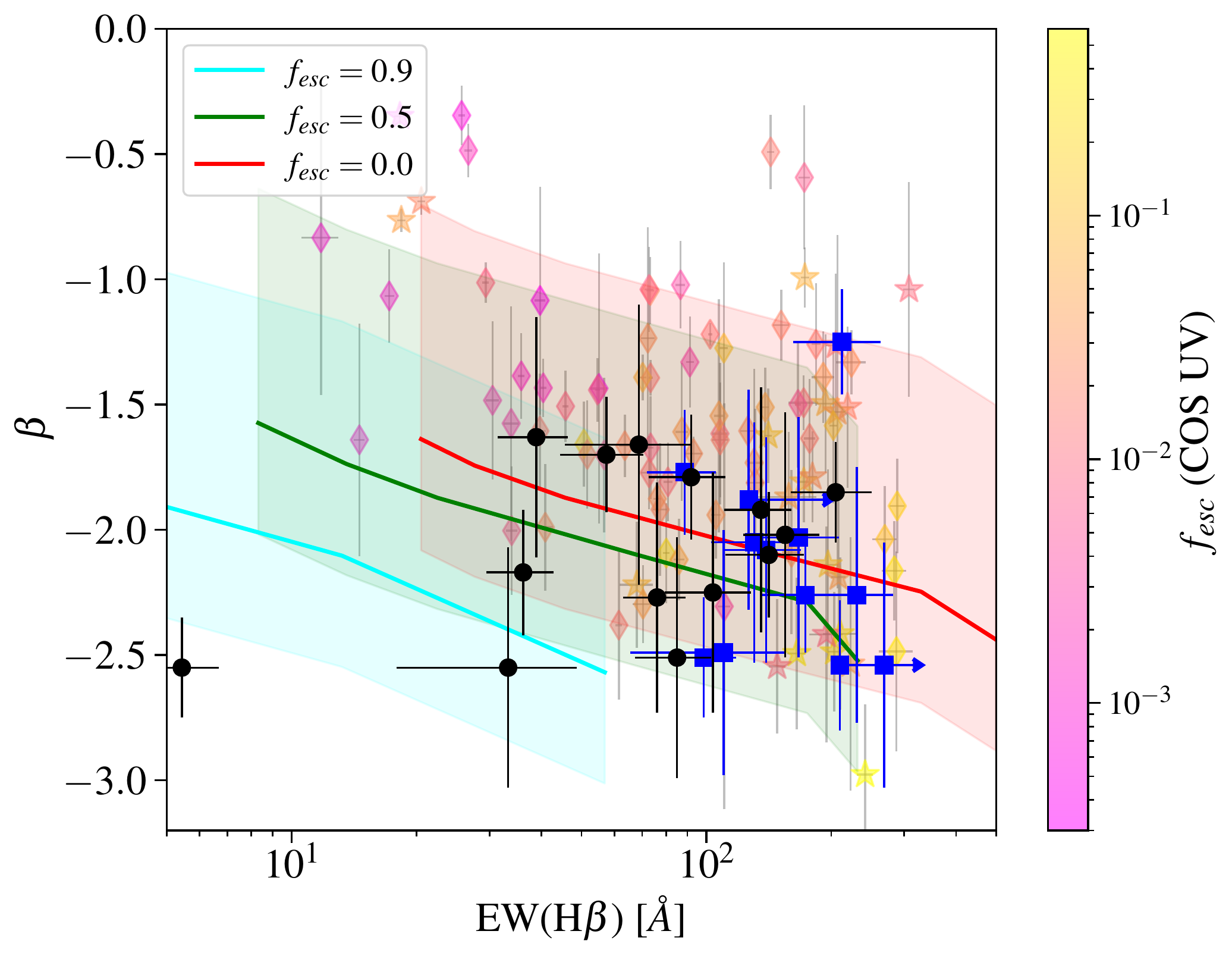}
\caption{$\beta$ vs. rest-frame $EW_0(\hb)$. Models are from \cite{Zackrisson2013} and simulate the expected trend for galaxies with an exponential declining SFR ($Z = 0.02$, solid lines) and various values of escape fractions. Symbols are as in Fig. \ref{fig:O32_Hbeta}.
\label{fig:beta_hb}}
\end{figure}

We show the properties of our sample in Fig. \ref{fig:beta_hb}. Specifically, at a given value of $EW_0(\hb)$, the high-redshift galaxies are, on average, bluer than the majority of the low-z galaxies (consistent with the lower stellar mass probed by our sources); but their slopes are similar to the subset of the low-z sources with moderate-to-high values for $f_{esc}$. The $\beta$ slopes for our galaxies are perfectly consistent with the average values found for LBGs at z$\sim 6$ for galaxies with $M_{UV}$ in the same range \cite{Bouwens2014}.
 In the figure, we also show the models from \cite{Zackrisson2013} for various values of escape fraction, after correcting the intrinsic $\beta$ slopes of the models for dust attenuation and assuming the average reddening of our sample derived from the SED fitting, $E(B-V) = 0.097$ (solid line), and also the maximum and minimum $E(B-V)$ in our sample (shaded area). We note that our galaxies show more consistency with model predictions that assume moderate-to-low escape fractions $0.0 \leq f_{esc} \leq 0.5$. 

Finally, we analyzed the sizes and star formation rate density for our galaxies. Several authors have postulated that concentrated star formation provides the feedback necessary to clear paths in the ISM that allow for the escape of LyC photons. In addition,  \cite{Marchi_2018} recently found that stacks of galaxies that are UV-compact ($r_{UV} \leq 0.30$ kpc) have much higher LyC flux than the average population 
\citep[see also][]{Izotov2018b}.
We find that, with the exception of one galaxy (ID 160281, which has $r_e =1.65 \text{kpc}$), our targets are all extremely compact with typical $r_e$ in the rest-frame UV around $0.2-0.6$ kpc.
These values are similar or even lower than those found for the low-z galaxies and LCEs \citep{Flury2022}; thus, once again the high-redshift sources have equal properties to the low-z LCEs.
Since we know that sizes evolve with redshift and galaxies become progressively more compact, we also compared our sample to the general population of star-forming galaxies (LBGs selected) at $z\simeq 6$ analyzed by \cite{Shibuya2015}. Specifically, for $M_{UV} =-18 (-20)$, the average UV rest-frame sizes are $r_e \simeq 0.38(0.56)$ kpc, respectively. Thus, indeed our galaxies are, from this point of view, as compact as the general UV faint population at the same redshift. 

As for the star formation rate densities,  the values of $\log_{10}(\Sigma_{SFR})$ for our galaxies span a very wide range, with an average $\Sigma_{SFR}$ that is slightly  higher than the average expected at their redshift, as measured by \cite{Shibuya2015}. We note that their values are inferred  from the UV and then dust corrected. LyC leakers at low and intermediate redshift, tend to show high  $\Sigma_{SFR}$  values, as discussed extensively by \cite{Naidu2020}, who actually proposed a physically motivated model in which $f_{esc}$ would scale as $\Sigma_{SFR}^{0.4}$. \cite{Flury2022} identify a threshold value of $\Sigma_{SFR} = 10 \ M_\odot \ \text{yr}^{-1} \ \text{kpc}^{-2}$ above which their LCE fraction changes from 10\% to 60\%, and where indeed we find half of our high-redshift galaxies.

\subsection{O32-R23 diagnostics}

\begin{figure*}[ht!]
\centering
\includegraphics[width=0.9\textwidth]{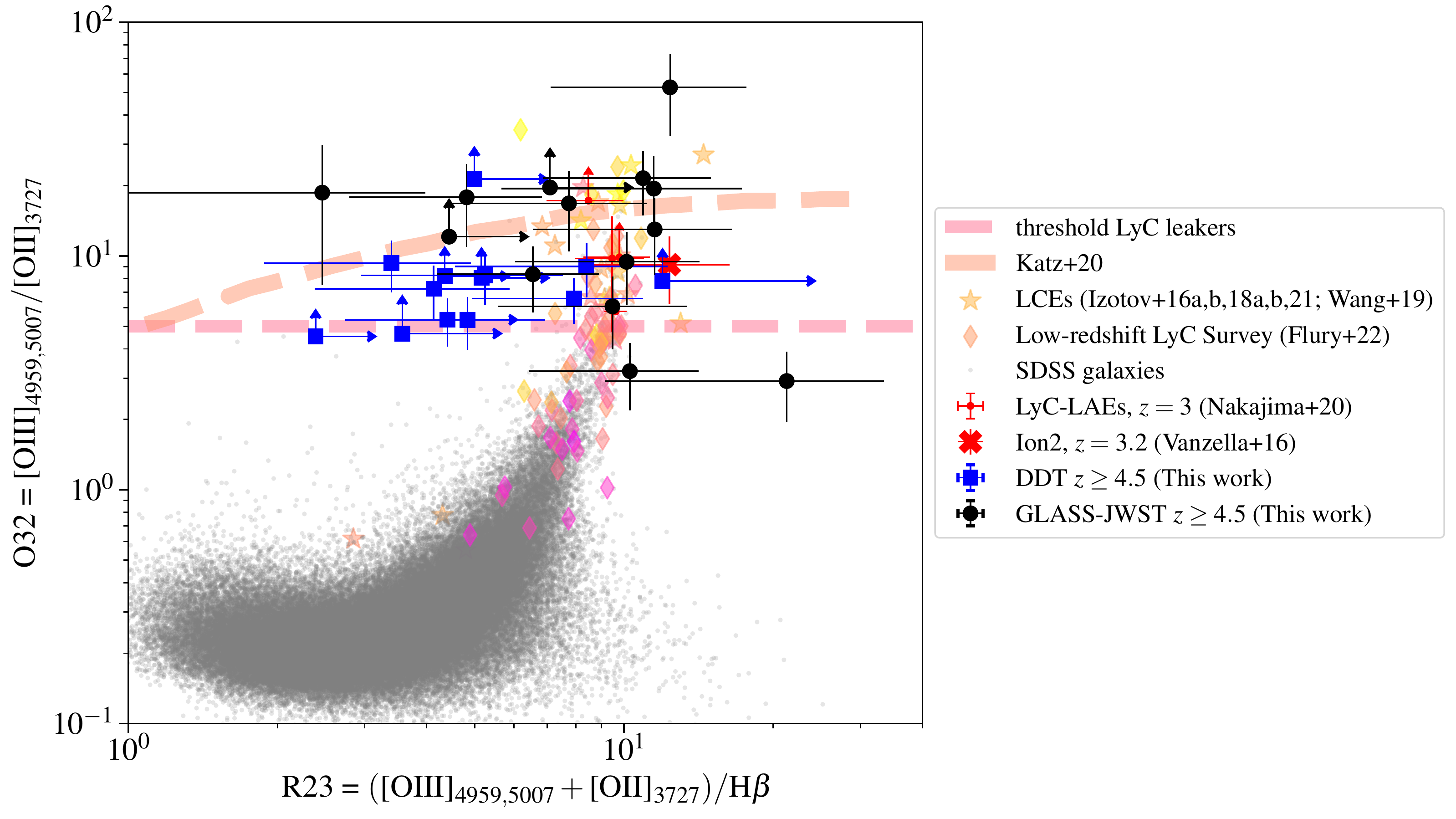}
\caption{$O32$ vs. $R23$ diagnostic diagram for our sample. Black dots and green squares show the GLASS-JWST and DDT samples, respectively. For reference, we plot also the SDSS galaxies from \cite{Thomas2013} as dark grey points. The LCEs from \cite{Nakajima2020}  are shown red dots. The Ion2 at $z = 3.2$ comes from \cite{Vanzella2016, deBarros2016} as a red x sign, the low-z LyC leakers from \cite{Izotov2016a, Izotov2016b, Izotov2018a, Izotov2018b, Wang2019} as stars, and the LCEs at $z = 0.3 - 0.4$ from the Low-redshift Lyman Continuum Survey \citep[as pink diamonds,][]{Flury2022}. We show also the locus of high-redshift LCEs predicted from cosmological simulations by \cite{Katz2020} as an orange dashed line and the threshold of $O32 > 5$ determined in \cite{Flury2022} as a pink dashed line.
\label{fig:O32_R23}}
\end{figure*}

The $O32$ vs $R23$ index diagram (Fig. \ref{fig:O32_R23}) is widely used to examine the gas-phase metallicity and ionization state both in the local universe \citep[e.g.,][]{Thomas2013, Izotov2016a, Izotov2016b, Izotov2018a, Izotov2018b} and at high redshift \citep[e.g.,][]{Flury2022, Nakajima2020, Reddy2022, Vanzella2019}, as both these indices are sensitive to combination of these quantities. \cite{Nakajima2020} showed that $z\sim 3$ LyC leakers tend to populate the upper right part of this diagram.
Recently \cite{Katz2020} used high-resolution cosmological radiation hydrodynamics simulations to examine the properties of LyC leakers deep into the epoch of reionization and found that simulated high-redshift galaxies populate the same regions of the $R23-O32$ plane, as the z$\sim$ 3 LyC leakers presented in \cite{Nakajima2020} that tend to have low metallicity. 
Although they conclude that this plane is not the most useful to differentiate between leakers and non-leakers we note that the $z=3$ leakers by \cite{Nakajima2020}, with measured values and Ion2 occupy the same region as the $z=0.3$ low-redshift leakers. Our galaxies occupy a much broader region, which could reflect either a wider range of metallicity and ionization states or simply the fact that we have large measured uncertainties on the diagnostics.

\section{Predicting escape fractions of EoR galaxies}\label{sec:predictions}

In the sections above, we have shown that our high-redshift galaxies have properties that are largely overlapping with those of low-z LCEs. The next step is to try and give an indirect estimate of the $f_{esc}$ values for our galaxies, so as to understand if typical low-mass galaxies at $z \simeq 5-7$ could be really the drivers of reionization. We assume that the mechanisms that drive the escape of LyC photons are the same at all redshifts and depend only on the physical properties of the sources. 

We proceeded as follows: we used the spectroscopic and physical properties of the 66 galaxies that are part of the \cite{Flury2022} sample, with the additional 22 LCEs from previous studies \citep{Izotov2016a, Izotov2016b, Izotov2018a, Izotov2018b, Izotov2021, Wang2019} to calibrate an empirical relation with the measured $f_{esc}$ values. Specifically, we focused on the following properties: $O32$, $EW_0(\hb)$, $\beta$, $r_e$ (in kpc), $\Sigma_{SFR}$, and $M_{1500}$. For all 88 galaxies, these parameters are accurately measured and, of course, accurate measurements of $f_{esc}$ are available. To this end, we  specifically used the $f_{esc}$ derived by the COS UV spectral fits \citep[see definition in][]{Flury2022}. 
Recently \cite{chisholm2022} followed a somewhat similar approach, also using the same set of low-redshift observations,  but limited to the UV-$\beta$ slope and an indirect proxy. They  provided  a scaling relation between $\beta$ and $f_{esc}$, although the relation has appreciable scatter that scales with $f_{esc}$. 

 We first ran the Spearman rank correlation and found that $O32$, $r_e$, $EW_0(\hb)$, and $\beta$ (in that order) are the properties that are best correlated with $f_{esc}$. We then identified one useful fitting linear relation to obtain an estimate of $\log_{10}(f_{esc})$ on the basis of the other four measured physical properties. A fully data-driven regression analysis was carried out by performing a regularized minimization of the root-mean-square error (MSE), computed between the values provided by the equation and the dataset, for several possible combinations of the above properties.
We tested both a scheme in which the error on the escape fraction, $f_{esc}$, is not considered and a scheme in which values are weighted by the error.
Since $O32$ and $EW_0(\hb)$ exhibit a very tight correlation (Spearman correlation between them $> 0.9$, see also Fig. \ref{fig:O32_Hbeta}), the information they provide is redundant and therefore we decided to use only $O32$. We checked that the remaining three parameters are reasonably independent (Spearman correlation $<$ 0.5) and therefore provide complementary information. We finally found a best-fit relation of the form: 
\begin{equation}
\log_{10}(f_{esc}) =A+B \log_{10}(O32)+ C r_e + D \beta.
\end{equation}
 After identifying the best type of equation, we repeated the minimization process 100 times, following a bootstrap approach every time a random number of sources (between 1 and 25, to avoid being left with too few sources) is randomly removed from the sample. In this way, we could constrain the confidence interval for the above best fit parameters. We find $A = -1.92 [-2.51, -1.71]$, $B = 0.48 [0.38, 0.69]$, $C = -0.96 [-1.20, -0.62]$, $D = -0.41 [-0.58, -0.31]$, where the values between the parentheses are in the 95th percentile distribution.

Testing the relation on the residuals, we find that it tends to slightly overestimates the $f_{esc}$ at very low values (much lower than 0.01), while it tends to underestimate the $f_{esc}$ at values that are higher than $\sim$0.1-0.2. This is due to the fact that in the sample used to fit the relation, there are very few galaxies with high $f_{esc}$ values and, in general, their measured $f_{esc}$ values have higher errors. 

\begin{figure}[ht!]
\includegraphics[width=\linewidth]{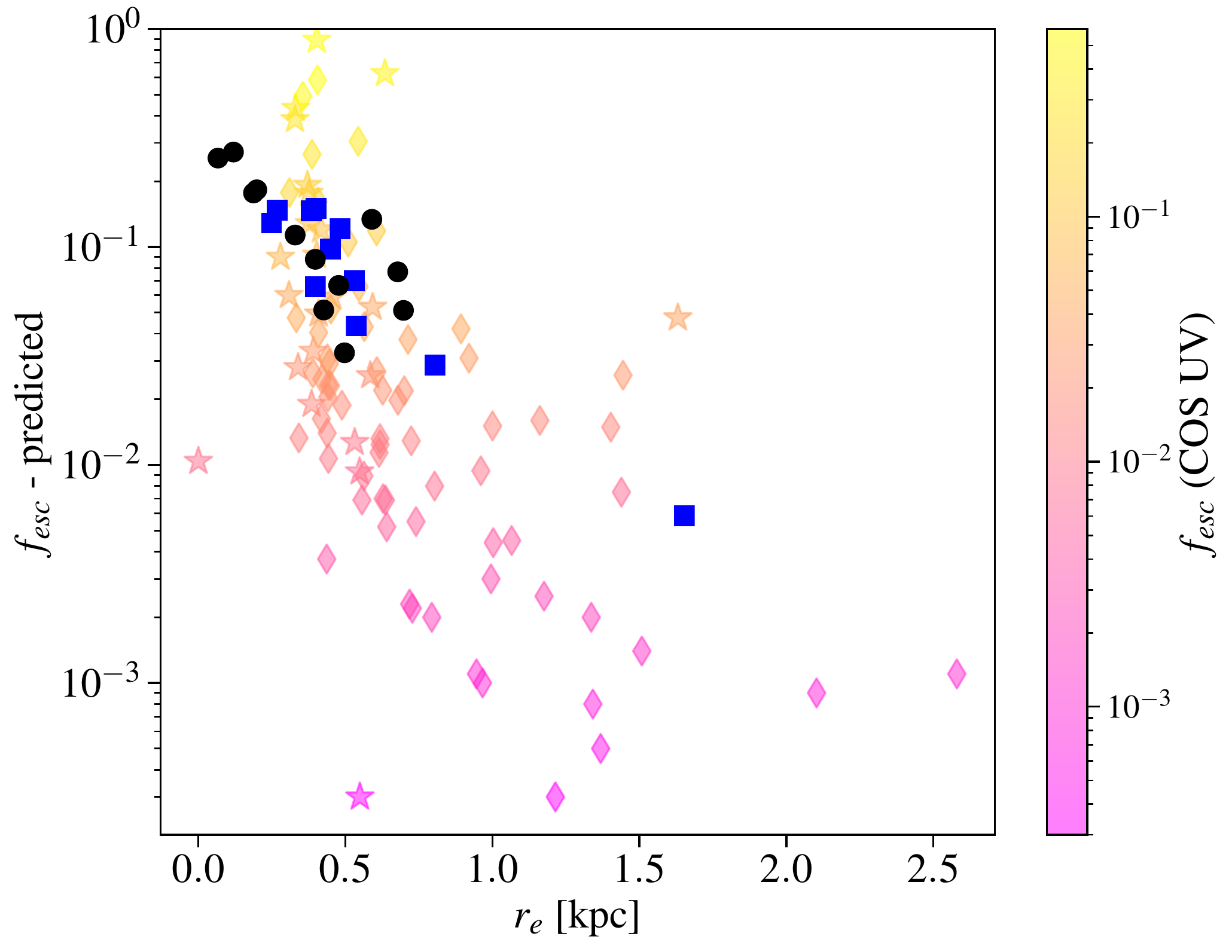}
\caption{Predicted $f_{esc}$ vs. size $r_e$ for our sample (blue and black symbols) and measured $f_{esc}$ vs. size $r_e$ from the literature. Symbols are as in Fig. \ref{fig:O32_Hbeta}. 
\label{fig:fesc_re}}
\end{figure}

We finally applied the above relation to our sample of high-redshift galaxies: out of 29 sources, 3 galaxies do not have the $r_e$ measurement since they are outside the UNCOVER footprint; for another 2, we do not have an estimate for the $O32$ parameter. We therefore could apply the best-fitting relation only to 24 sources. In Fig. \ref{fig:fesc_re}, we show the predicted $f_{esc}$ for our sources as a function of $r_e$, as well as the low-redshift comparison sample (for which the $f_{esc}$ are measured values). Most of our galaxies have predicted $f_{esc}$ values larger than 0.05, meaning that they would be considered leakers. The average $f_{esc}$ is $\simeq 0.12$ with the bluest and most compact sources having $f_{esc}$ as large as $0.2-0.4$, which could be lower limits for the reasons discussed above. 

Clearly, the main limitations of the above analysis are the fact that the low-z sample used to calibrate the relation is small and, most importantly, it is not evenly populated as it contains mostly objects with rather low $f_{esc}$.
Also, it would be important to test these predictions at intermediate redshift ($z \sim 3-4$), that is, much closer to the EoR, where it is possible to both directly detect LyC emission and determine all other physical and spectroscopic properties.
However, at the moment, measurements of  $f_{esc}$ at intermediate redshift are still sparse and most samples lack near-infrared (NIR) spectroscopic follow-up results. 
In spite of these limitations, a consistent picture emerges from our results, suggesting low-mass galaxies at $z\sim5-7$ have mostly properties which indicate moderate values of $f_{esc}$. Interestingly, the average $f_{esc}$ value inferred for our sample is equal to the one predicted by \cite{Naidu2020} for galaxies in the mass range $\log_{10}M_\star = 8-9$ at $z \sim 6$, according to their simple model, where $f_{esc}$ scales with $\Sigma_{SFR}$, which was constrained to reproduce the average observed values for the \cite{Steidel2018} sample at $z=3$.
%With assuming an average  $\xi_{ion}$), and 
%such galaxies would contribute to about 30-40\% of the total ionizing budget at that redshift, with the rest being dominated by even brighter/more massive sources ($\log_{10}M_\star \geq 9$). With this model, \cite{Naidu2020} are able to reproduce the late and rapid reionization scenario implied by the latest observational constrains on reionization, from Thomson optical depth \citep{Planck2020}, to \lya\ fraction in LBGs \citep{Pentericci2018,Mason2018a}, to QSOs damping wing \citep{Davies2019} and so on.

\section{Summary and conclusions}\label{sec:results}

Thanks to the magnification power of the Abell 2744 cluster, in this paper, we present the first \jwst/\NIRSpec\ observations of 29 gravitationally lensed galaxies with photometric or spectroscopic redshift in the range $4.5 \leq z \leq 8$. From a combined \NIRSpec\ and \NIRCam\ analysis, we were able to derive accurate physical properties of the galaxies, including $M_{\star}$, SFR, $r_e$, UV-$\beta$ slopes, and measurements of the most prominent optical emission lines (\oii, \oiiialone, \hb, \ha). We summarize our findings as follows:

\begin{itemize}

\item Our sample is composed of purely star-forming galaxies, as inferred from the MEx diagram that excludes the presence of any AGN. 
\item The galaxies in our sample have blue UV slopes (median $\beta$ = -2.08) and are mostly very compact with $r_e \simeq 0.2-0.5$ kpc. These properties are consistent with those of the general population of LBGs at similar redshift and with similarly faint $M_{UV}$ \citep{Bouwens2014,Shibuya2015}.
 \item Compared to the low-z sample by \cite{Flury2022} as well as to the low-z LCEs from previous studies, our galaxies present properties (in terms of $O32$, $EW_0(\hb)$, UV-$\beta$ slopes, $r_e$, and $\Sigma_{SFR}$) that are entirely consistent with those of low-z galaxies with measured $f_{esc}$ larger than 0.05 and which are considered to be LyC leakers.
 
\item Using a linear analysis that employs the minimization of MSE, we found a best fitting relation between $f_{esc}$ and the three most correlated and independent parameters ($O32$, UV-$\beta$, and $r_e$) for the low-redshift sample of 88 galaxies. Applying this relation to the 24 galaxies from our sample where these parameters are all measured, we find that 20/24 have predicted escape fractions larger than 0.05, that is, they would be considered leakers and the average $f_{esc}$ of our sample is 0.12.
\end{itemize}

In conclusion, our results show that indeed the average low-mass galaxies around the epoch of reionization have physical and spectroscopic properties consistent with moderate values of escaping ionizing photons ($f_{esc}= 0.1-0.2$). 
With upcoming \jwst\ observations from GTO and GO programs, this analysis could be easily extended to larger samples of galaxies with higher masses and/or at higher redshift to determine if the bulk of the reionizing photons have been provided by even more massive and brighter galaxies \citep[as predicted, e.g., by][]{Sharma2016,Naidu2020} or if we need a more substantial contribution by the fainter and more numerous galaxies with $M_{UV} > -18,$ as predicted by most other popular models \citep[e.g.,][]{Trebitsch2022,Finkelstein2019, Atek2015}.

\begin{acknowledgements}

Support for program JWST-ERS-1324 was provided by NASA through a grant from the Space Telescope Science Institute, which is operated by the Association of Universities for Research in Astronomy, Inc., under NASA contract NAS 5-03127.

This work is based on observations made with the NASA/ESA/CSA James Webb Space Telescope. The data were obtained from the Mikulski Archive for Space Telescopes at the Space Telescope Science Institute, which is operated by the Association of Universities for Research in Astronomy, Inc., under NASA contract NAS 5-03127 for JWST. These observations are associated with programs GLASS-JWST \#1324, JWST DDT \#2756 and GO UNCOVER \#2561.

SM thanks the Department of Physics and Astronomy at the University of California, Los Angeles, for a productive and satisfying visit in which the majority of the work presented in this paper was done.

We acknowledge support from the INAF Large Grant 2022 “Extragalactic Surveys with JWST” (PI Pentericci). 

This research is supported in part by the Australian Research Council Centre of Excellence for All Sky Astrophysics in 3 Dimensions (ASTRO 3D), through project number CE170100013.

LY acknowledges support by JSPS KAKENHI Grant Number JP 21F21325. RAW acknowledges support from NASA JWST Interdisciplinary Scientist grants NAG5-12460, NNX14AN10G and 80NSSC18K0200 from GSFC.

MB acknowledges support from the Slovenian national research agency ARRS through grant N1-0238. 

CM acknowledges support by the VILLUM FONDEN under grant 37459. The Cosmic Dawn Center (DAWN) is funded by the Danish National Research Foundation under grant DNRF140.
KG and TN acknowledge support from Australian Research Council Laureate Fellowship FL180100060.
\end{acknowledgements}

%\section{Appendix information}

\bibliographystyle{aa}
\bibliography{biblio}

\begin{thebibliography}{86}
\expandafter\ifx\csname natexlab\endcsname\relax\def\natexlab#1{#1}\fi

\bibitem[{{Alavi} {et~al.}(2020){Alavi}, {Colbert}, {Teplitz}, {Siana},
  {Scarlata}, {Rutkowski}, {Mehta}, {Henry}, {Dai}, {Haardt}, \&
  {Bagley}}]{Alavi2020}
{Alavi}, A., {Colbert}, J., {Teplitz}, H.~I., {et~al.} 2020, \apj, 904, 59

\bibitem[{{Atek} {et~al.}(2015){Atek}, {Richard}, {Kneib}, {Jauzac},
  {Schaerer}, {Clement}, {Limousin}, {Jullo}, {Natarajan}, {Egami}, \&
  {Ebeling}}]{Atek2015}
{Atek}, H., {Richard}, J., {Kneib}, J.-P., {et~al.} 2015, \apj, 800, 18

\bibitem[{{Baldwin} {et~al.}(1981){Baldwin}, {Phillips}, \&
  {Terlevich}}]{BPT_1981}
{Baldwin}, J.~A., {Phillips}, M.~M., \& {Terlevich}, R. 1981, \pasp, 93, 5

\bibitem[{{Bassett} {et~al.}(2019){Bassett}, {Ryan-Weber}, {Cooke}, {Diaz},
  {Nanayakkara}, {Yuan}, {Spitler}, {Me{\v{s}}tri{\'c}}, {Garel}, {Sawicki},
  {Gwyn}, \& {Golob}}]{Bassett2019}
{Bassett}, R., {Ryan-Weber}, E.~V., {Cooke}, J., {et~al.} 2019, \mnras, 483,
  5223

\bibitem[{{Bergamini} {et~al.}(2022){Bergamini}, {Acebron}, {Grillo}, {Rosati},
  {Caminha}, {Mercurio}, {Vanzella}, {Angora}, {Brammer}, {Meneghetti}, \&
  {Nonino}}]{Bergamini2022}
{Bergamini}, P., {Acebron}, A., {Grillo}, C., {et~al.} 2022, arXiv e-prints,
  arXiv:2207.09416

\bibitem[{{Bergvall} {et~al.}(2013){Bergvall}, {Leitet}, {Zackrisson}, \&
  {Marquart}}]{bergvall2013}
{Bergvall}, N., {Leitet}, E., {Zackrisson}, E., \& {Marquart}, T. 2013, \aap,
  554, A38

\bibitem[{{Binggeli} {et~al.}(2018){Binggeli}, {Zackrisson}, {Pelckmans},
  {Cubo}, {Jensen}, \& {Shimizu}}]{Binggeli2018}
{Binggeli}, C., {Zackrisson}, E., {Pelckmans}, K., {et~al.} 2018, \mnras, 479,
  368

\bibitem[{{Birrer} {et~al.}(2021){Birrer}, {Shajib}, {Gilman}, {Galan},
  {Aalbers}, {Millon}, {Morgan}, {Pagano}, {Park}, {Teodori}, {Tessore},
  {Ueland}, {Van de Vyvere}, {Wagner-Carena}, {Wempe}, {Yang}, {Ding},
  {Schmidt}, {Sluse}, {Zhang}, \& {Amara}}]{Birrer2021}
{Birrer}, S., {Shajib}, A., {Gilman}, D., {et~al.} 2021, The Journal of Open
  Source Software, 6, 3283

\bibitem[{{Bolan} {et~al.}(2021){Bolan}, {Lemaux}, {Mason}, {Brada{\v{c}}},
  {Treu}, {Strait}, {Pelliccia}, {Pentericci}, \& {Malkan}}]{bolan2021}
{Bolan}, P., {Lemaux}, B.~C., {Mason}, C., {et~al.} 2021, arXiv e-prints,
  arXiv:2111.14912

\bibitem[{{Bouwens} {et~al.}(2014){Bouwens}, {Illingworth}, {Oesch},
  {Labb{\'e}}, {van Dokkum}, {Trenti}, {Franx}, {Smit}, {Gonzalez}, \&
  {Magee}}]{Bouwens2014}
{Bouwens}, R.~J., {Illingworth}, G.~D., {Oesch}, P.~A., {et~al.} 2014, \apj,
  793, 115

\bibitem[{{Bruzual} \& {Charlot}(2003)}]{Bruzual2003}
{Bruzual}, G. \& {Charlot}, S. 2003, \mnras, 344, 1000

\bibitem[{{Calabr{\`o}} {et~al.}(2021){Calabr{\`o}}, {Castellano},
  {Pentericci}, {Fontanot}, {Menci}, {Cullen}, {McLure}, {Bolzonella},
  {Cimatti}, {Marchi}, {Talia}, {Amor{\'\i}n}, {Cresci}, {De Lucia}, {Fynbo},
  {Fontana}, {Franco}, {Hathi}, {Hibon}, {Hirschmann}, {Mannucci}, {Santini},
  {Saxena}, {Schaerer}, {Xie}, \& {Zamorani}}]{Calabro2021}
{Calabr{\`o}}, A., {Castellano}, M., {Pentericci}, L., {et~al.} 2021, \aap,
  646, A39

\bibitem[{{Calzetti} {et~al.}(2000){Calzetti}, {Armus}, {Bohlin}, {Kinney},
  {Koornneef}, \& {Storchi-Bergmann}}]{Calzetti2000}
{Calzetti}, D., {Armus}, L., {Bohlin}, R.~C., {et~al.} 2000, \apj, 533, 682

\bibitem[{{Castellano} {et~al.}(2016){Castellano}, {Dayal}, {Pentericci},
  {Fontana}, {Hutter}, {Brammer}, {Merlin}, {Grazian}, {Pilo}, {Amorin},
  {Cristiani}, {Dickinson}, {Ferrara}, {Gallerani}, {Giallongo}, {Giavalisco},
  {Guaita}, {Koekemoer}, {Maiolino}, {Paris}, {Santini}, {Vallini}, {Vanzella},
  \& {Wagg}}]{Castellano2016}
{Castellano}, M., {Dayal}, P., {Pentericci}, L., {et~al.} 2016, \apjl, 818, L3

\bibitem[{{Chabrier}(2003)}]{Chabrier2003}
{Chabrier}, G. 2003, \pasp, 115, 763

\bibitem[{{Chisholm} {et~al.}(2020){Chisholm}, {Prochaska}, {Schaerer},
  {Gazagnes}, \& {Henry}}]{chisholm2020}
{Chisholm}, J., {Prochaska}, J.~X., {Schaerer}, D., {Gazagnes}, S., \& {Henry},
  A. 2020, \mnras, 498, 2554

\bibitem[{{Chisholm} {et~al.}(2022){Chisholm}, {Saldana-Lopez}, {Flury},
  {Schaerer}, {Jaskot}, {Amor{\'\i}n}, {Atek}, {Finkelstein}, {Fleming},
  {Ferguson}, {Fern{\'a}ndez}, {Giavalisco}, {Hayes}, {Heckman}, {Henry}, {Ji},
  {Marques-Chaves}, {Mauerhofer}, {McCandliss}, {Oey}, {{\"O}stlin},
  {Rutkowski}, {Scarlata}, {Thuan}, {Trebitsch}, {Wang}, {Worseck}, \&
  {Xu}}]{chisholm2022}
{Chisholm}, J., {Saldana-Lopez}, A., {Flury}, S., {et~al.} 2022, \mnras, 517,
  5104

\bibitem[{{Coil} {et~al.}(2015){Coil}, {Aird}, {Reddy}, {Shapley}, {Kriek},
  {Siana}, {Mobasher}, {Freeman}, {Price}, \& {Shivaei}}]{Coil2015}
{Coil}, A.~L., {Aird}, J., {Reddy}, N., {et~al.} 2015, \apj, 801, 35

\bibitem[{{Curti} {et~al.}(2023){Curti}, {D'Eugenio}, {Carniani}, {Maiolino},
  {Sandles}, {Witstok}, {Baker}, {Bennett}, {Piotrowska}, {Tacchella},
  {Charlot}, {Nakajima}, {Maheson}, {Mannucci}, {Amiri}, {Arribas}, {Belfiore},
  {Bonaventura}, {Bunker}, {Chevallard}, {Cresci}, {Curtis-Lake},
  {Hayden-Pawson}, {Jones}, {Kumari}, {Laseter}, {Looser}, {Marconi}, {Maseda},
  {Scholtz}, {Smit}, {{\"U}bler}, \& {Wallace}}]{curti2023}
{Curti}, M., {D'Eugenio}, F., {Carniani}, S., {et~al.} 2023, \mnras, 518, 425

\bibitem[{{de Barros} {et~al.}(2016){de Barros}, {Vanzella}, {Amor{\'{\i}}n},
  {Castellano}, {Siana}, {Grazian}, {Suh}, {Balestra}, {Vignali}, {Verhamme},
  {Zamorani}, {Mignoli}, {Hasinger}, {Comastri}, {Pentericci},
  {P{\'e}rez-Montero}, {Fontana}, {Giavalisco}, \& {Gilli}}]{deBarros2016}
{de Barros}, S., {Vanzella}, E., {Amor{\'{\i}}n}, R., {et~al.} 2016, \aap, 585,
  A51

\bibitem[{{Dom{\'\i}nguez} {et~al.}(2013){Dom{\'\i}nguez}, {Siana}, {Henry},
  {Scarlata}, {Bedregal}, {Malkan}, {Atek}, {Ross}, {Colbert}, {Teplitz},
  {Rafelski}, {McCarthy}, {Bunker}, {Hathi}, {Dressler}, {Martin}, \&
  {Masters}}]{Dominguez2013}
{Dom{\'\i}nguez}, A., {Siana}, B., {Henry}, A.~L., {et~al.} 2013, \apj, 763,
  145

\bibitem[{{Dopita} \& {Sutherland}(2003)}]{Dopita2003}
{Dopita}, M.~A. \& {Sutherland}, R.~S. 2003, {Astrophysics of the diffuse
  universe}

\bibitem[{{Faisst} {et~al.}(2018){Faisst}, {Masters}, {Wang}, {Merson},
  {Capak}, {Malhotra}, \& {Rhoads}}]{Faisst2018}
{Faisst}, A.~L., {Masters}, D., {Wang}, Y., {et~al.} 2018, \apj, 855, 132

\bibitem[{{Finkelstein} {et~al.}(2019){Finkelstein}, {D'Aloisio},
  {Paardekooper}, {Ryan}, {Behroozi}, {Finlator}, {Livermore}, {Upton
  Sanderbeck}, {Dalla Vecchia}, \& {Khochfar}}]{Finkelstein2019}
{Finkelstein}, S.~L., {D'Aloisio}, A., {Paardekooper}, J.-P., {et~al.} 2019,
  \apj, 879, 36

\bibitem[{{Flury} {et~al.}(2022){Flury}, {Jaskot}, {Ferguson}, {Worseck},
  {Makan}, {Chisholm}, {Saldana-Lopez}, {Schaerer}, {McCandliss}, {Wang},
  {Ford}, {Heckman}, {Ji}, {Giavalisco}, {Amorin}, {Atek}, {Blaizot},
  {Borthakur}, {Carr}, {Castellano}, {Cristiani}, {De Barros}, {Dickinson},
  {Finkelstein}, {Fleming}, {Fontanot}, {Garel}, {Grazian}, {Hayes}, {Henry},
  {Mauerhofer}, {Micheva}, {Oey}, {Ostlin}, {Papovich}, {Pentericci},
  {Ravindranath}, {Rosdahl}, {Rutkowski}, {Santini}, {Scarlata}, {Teplitz},
  {Thuan}, {Trebitsch}, {Vanzella}, {Verhamme}, \& {Xu}}]{Flury2022}
{Flury}, S.~R., {Jaskot}, A.~E., {Ferguson}, H.~C., {et~al.} 2022, \apjs, 260,
  1

\bibitem[{{Fontana} {et~al.}(2000){Fontana}, {D'Odorico}, {Poli}, {Giallongo},
  {Arnouts}, {Cristiani}, {Moorwood}, \& {Saracco}}]{fontana00}
{Fontana}, A., {D'Odorico}, S., {Poli}, F., {et~al.} 2000, \aj, 120, 2206

\bibitem[{{Gazagnes} {et~al.}(2020){Gazagnes}, {Chisholm}, {Schaerer},
  {Verhamme}, \& {Izotov}}]{Gazagnes2020}
{Gazagnes}, S., {Chisholm}, J., {Schaerer}, D., {Verhamme}, A., \& {Izotov}, Y.
  2020, \aap, 639, A85

\bibitem[{{Inoue} {et~al.}(2014){Inoue}, {Shimizu}, {Iwata}, \&
  {Tanaka}}]{inoue2014}
{Inoue}, A.~K., {Shimizu}, I., {Iwata}, I., \& {Tanaka}, M. 2014, \mnras, 442,
  1805

\bibitem[{{Izotov} {et~al.}(2022){Izotov}, {Chisholm}, {Worseck}, {Guseva},
  {Schaerer}, \& {Prochaska}}]{izotov2022}
{Izotov}, Y.~I., {Chisholm}, J., {Worseck}, G., {et~al.} 2022, \mnras, 515,
  2864

\bibitem[{{Izotov} {et~al.}(2021){Izotov}, {Guseva}, {Fricke}, {Henkel},
  {Schaerer}, \& {Thuan}}]{Izotov2021}
{Izotov}, Y.~I., {Guseva}, N.~G., {Fricke}, K.~J., {et~al.} 2021, \aap, 646,
  A138

\bibitem[{{Izotov} {et~al.}(2016{\natexlab{a}}){Izotov}, {Orlitov{\'a}},
  {Schaerer}, {Thuan}, {Verhamme}, {Guseva}, \& {Worseck}}]{Izotov2016a}
{Izotov}, Y.~I., {Orlitov{\'a}}, I., {Schaerer}, D., {et~al.}
  2016{\natexlab{a}}, \nat, 529, 178

\bibitem[{{Izotov} {et~al.}(2016{\natexlab{b}}){Izotov}, {Schaerer}, {Thuan},
  {Worseck}, {Guseva}, {Orlitov{\'a}}, \& {Verhamme}}]{Izotov2016b}
{Izotov}, Y.~I., {Schaerer}, D., {Thuan}, T.~X., {et~al.} 2016{\natexlab{b}},
  \mnras, 461, 3683

\bibitem[{{Izotov} {et~al.}(2018{\natexlab{a}}){Izotov}, {Schaerer}, {Worseck},
  {Guseva}, {Thuan}, {Verhamme}, {Orlitov{\'a}}, \& {Fricke}}]{Izotov2018a}
{Izotov}, Y.~I., {Schaerer}, D., {Worseck}, G., {et~al.} 2018{\natexlab{a}},
  \mnras, 474, 4514

\bibitem[{{Izotov} {et~al.}(2018{\natexlab{b}}){Izotov}, {Worseck}, {Schaerer},
  {Guseva}, {Thuan}, {Fricke}, \& {Orlitov{\'a}}}]{Izotov2018b}
{Izotov}, Y.~I., {Worseck}, G., {Schaerer}, D., {et~al.} 2018{\natexlab{b}},
  \mnras, 478, 4851

\bibitem[{{Juneau} {et~al.}(2014){Juneau}, {Bournaud}, {Charlot}, {Daddi},
  {Elbaz}, {Trump}, {Brinchmann}, {Dickinson}, {Duc}, {Gobat}, {Jean-Baptiste},
  {Le Floc'h}, {Lehnert}, {Pacifici}, {Pannella}, \& {Schreiber}}]{Juneau2014}
{Juneau}, S., {Bournaud}, F., {Charlot}, S., {et~al.} 2014, \apj, 788, 88

\bibitem[{{Juneau} {et~al.}(2011){Juneau}, {Dickinson}, {Alexander}, \&
  {Salim}}]{Juneau2011}
{Juneau}, S., {Dickinson}, M., {Alexander}, D.~M., \& {Salim}, S. 2011, \apj,
  736, 104

\bibitem[{{Jung} {et~al.}(2020){Jung}, {Finkelstein}, {Dickinson}, {Hutchison},
  {Larson}, {Papovich}, {Pentericci}, {Straughn}, {Guo}, {Malhotra}, {Rhoads},
  {Song}, {Tilvi}, \& {Wold}}]{jung2020}
{Jung}, I., {Finkelstein}, S.~L., {Dickinson}, M., {et~al.} 2020, \apj, 904,
  144

\bibitem[{{Kashino} {et~al.}(2013){Kashino}, {Silverman}, {Rodighiero},
  {Renzini}, {Arimoto}, {Daddi}, {Lilly}, {Sanders}, {Kartaltepe}, {Zahid},
  {Nagao}, {Sugiyama}, {Capak}, {Carollo}, {Chu}, {Hasinger}, {Ilbert},
  {Kajisawa}, {Kewley}, {Koekemoer}, {Kova{\v{c}}}, {Le F{\`e}vre}, {Masters},
  {McCracken}, {Onodera}, {Scoville}, {Strazzullo}, {Symeonidis}, \&
  {Taniguchi}}]{Kashino2013}
{Kashino}, D., {Silverman}, J.~D., {Rodighiero}, G., {et~al.} 2013, \apjl, 777,
  L8

\bibitem[{{Katz} {et~al.}(2020){Katz}, {{\v{D}}urov{\v{c}}{\'\i}kov{\'a}},
  {Kimm}, {Rosdahl}, {Blaizot}, {Haehnelt}, {Devriendt}, {Slyz}, {Ellis}, \&
  {Laporte}}]{Katz2020}
{Katz}, H., {{\v{D}}urov{\v{c}}{\'\i}kov{\'a}}, D., {Kimm}, T., {et~al.} 2020,
  \mnras, 498, 164

\bibitem[{{Kennicutt}(1998)}]{Kennicutt1998}
{Kennicutt}, Jr., R.~C. 1998, \araa, 36, 189

\bibitem[{{Mahler} {et~al.}(2018){Mahler}, {Richard}, {Cl{\'e}ment},
  {Lagattuta}, {Schmidt}, {Patr{\'\i}cio}, {Soucail}, {Bacon}, {Pello},
  {Bouwens}, {Maseda}, {Martinez}, {Carollo}, {Inami}, {Leclercq}, \&
  {Wisotzki}}]{mahler2018}
{Mahler}, G., {Richard}, J., {Cl{\'e}ment}, B., {et~al.} 2018, \mnras, 473, 663

\bibitem[{{Malkan} \& {Malkan}(2021)}]{Malkan2021}
{Malkan}, M.~A. \& {Malkan}, B.~K. 2021, \apj, 909, 92

\bibitem[{{Marchi} {et~al.}(2018){Marchi}, {Pentericci}, {Guaita}, {Schaerer},
  {Verhamme}, {Castellano}, {Ribeiro}, {Garilli}, {Le F{\`e}vre}, {Amorin},
  {Bardelli}, {Cassata}, {Durkalec}, {Grazian}, {Hathi}, {Lemaux}, {Maccagni},
  {Vanzella}, \& {Zucca}}]{Marchi_2018}
{Marchi}, F., {Pentericci}, L., {Guaita}, L., {et~al.} 2018, \aap, 614, A11

\bibitem[{{Mascia} {et~al.}(2023){Mascia}, {Pentericci}, {Saxena}, {Belfiori},
  {Calabr{\`o}}, {Castellano}, {Saldana-Lopez}, {Talia}, {Amor{\'\i}n},
  {Cullen}, {Garilli}, {Guaita}, {Llerena}, {McLure}, {Moresco}, {Santini}, \&
  {Schaerer}}]{Mascia2023}
{Mascia}, S., {Pentericci}, L., {Saxena}, A., {et~al.} 2023, arXiv e-prints,
  arXiv:2301.09328

\bibitem[{{Mason} {et~al.}(2019){Mason}, {Fontana}, {Treu}, {Schmidt}, {Hoag},
  {Abramson}, {Amorin}, {Brada{\v{c}}}, {Guaita}, {Jones}, {Henry}, {Malkan},
  {Pentericci}, {Trenti}, \& {Vanzella}}]{mason2019}
{Mason}, C.~A., {Fontana}, A., {Treu}, T., {et~al.} 2019, \mnras, 485, 3947

\bibitem[{{Mason} {et~al.}(2018){Mason}, {Treu}, {Dijkstra}, {Mesinger},
  {Trenti}, {Pentericci}, {de Barros}, \& {Vanzella}}]{Mason2018a}
{Mason}, C.~A., {Treu}, T., {Dijkstra}, M., {et~al.} 2018, \apj, 856, 2

\bibitem[{{Merlin} {et~al.}(2022){Merlin}, {Bonchi}, {Paris}, {Belfiori},
  {Fontana}, {Castellano}, {Nonino}, {Polenta}, {Santini}, {Yang},
  {Glazebrook}, {Treu}, {Roberts-Borsani}, {Trenti}, {Birrer}, {Brammer},
  {Grillo}, {Calabr{\`o}}, {Marchesini}, {Mason}, {Mercurio}, {Morishita},
  {Strait}, {Boyett}, {Leethochawalit}, {Nanayakkara}, {Vulcani}, {Bradac}, \&
  {Wang}}]{Merlin2022}
{Merlin}, E., {Bonchi}, A., {Paris}, D., {et~al.} 2022, arXiv e-prints,
  arXiv:2207.11701

\bibitem[{{Morishita} {et~al.}(2022){Morishita}, {Roberts-Borsani}, {Treu},
  {Brammer}, {Mason}, {Trenti}, {Vulcani}, {Wang}, {Acebron}, {Bah{\'e}},
  {Bergamini}, {Boyett}, {Bradac}, {Calabr{\`o}}, {Castellano}, {Chen}, {De
  Lucia}, {Filippenko}, {Fontana}, {Glazebrook}, {Grillo}, {Henry}, {Jones},
  {Kelly}, {Koekemoer}, {Leethochawalit}, {Lu}, {Marchesini}, {Mascia},
  {Mercurio}, {Merlin}, {Metha}, {Nanayakkara}, {Nonino}, {Paris},
  {Pentericci}, {Santini}, {Strait}, {Vanzella}, {Windhorst}, {Rosati}, \&
  {Xie}}]{Morishita2022}
{Morishita}, T., {Roberts-Borsani}, G., {Treu}, T., {et~al.} 2022, arXiv
  e-prints, arXiv:2211.09097

\bibitem[{{Naidu} {et~al.}(2020){Naidu}, {Tacchella}, {Mason}, {Bose}, {Oesch},
  \& {Conroy}}]{Naidu2020}
{Naidu}, R.~P., {Tacchella}, S., {Mason}, C.~A., {et~al.} 2020, \apj, 892, 109

\bibitem[{{Nakajima} {et~al.}(2020){Nakajima}, {Ellis}, {Robertson}, {Tang}, \&
  {Stark}}]{Nakajima2020}
{Nakajima}, K., {Ellis}, R.~S., {Robertson}, B.~E., {Tang}, M., \& {Stark},
  D.~P. 2020, \apj, 889, 161

\bibitem[{{Nakajima} \& {Ouchi}(2014)}]{Nakajima2014}
{Nakajima}, K. \& {Ouchi}, M. 2014, \mnras, 442, 900

\bibitem[{{Nakajima} {et~al.}(2023){Nakajima}, {Ouchi}, {Isobe}, {Harikane},
  {Zhang}, {Ono}, {Umeda}, \& {Oguri}}]{nakajima2023}
{Nakajima}, K., {Ouchi}, M., {Isobe}, Y., {et~al.} 2023, arXiv e-prints,
  arXiv:2301.12825

\bibitem[{{Oke} \& {Gunn}(1983)}]{Oke1983}
{Oke}, J.~B. \& {Gunn}, J.~E. 1983, \apj, 266, 713

\bibitem[{{Ouchi} {et~al.}(2020){Ouchi}, {Ono}, \& {Shibuya}}]{Ouchi_2020}
{Ouchi}, M., {Ono}, Y., \& {Shibuya}, T. 2020, \araa, 58, 617

\bibitem[{{Pahl} {et~al.}(2021){Pahl}, {Shapley}, {Steidel}, {Chen}, \&
  {Reddy}}]{Pahl2021}
{Pahl}, A.~J., {Shapley}, A., {Steidel}, C.~C., {Chen}, Y., \& {Reddy}, N.~A.
  2021, \mnras, 505, 2447

\bibitem[{{Pentericci} {et~al.}(2011){Pentericci}, {Fontana}, {Vanzella},
  {Castellano}, {Grazian}, {Dijkstra}, {Boutsia}, {Cristiani}, {Dickinson},
  {Giallongo}, {Giavalisco}, {Maiolino}, {Moorwood}, {Paris}, \&
  {Santini}}]{Pentericci2011}
{Pentericci}, L., {Fontana}, A., {Vanzella}, E., {et~al.} 2011, \apj, 743, 132

\bibitem[{{Pentericci} {et~al.}(2018){Pentericci}, {Vanzella}, {Castellano},
  {Fontana}, {De Barros}, {Grazian}, {Marchi}, {Bradac}, {Conselice},
  {Cristiani}, {Dickinson}, {Finkelstein}, {Giallongo}, {Guaita}, {Koekemoer},
  {Maiolino}, {Santini}, \& {Tilvi}}]{Pentericci_2018b}
{Pentericci}, L., {Vanzella}, E., {Castellano}, M., {et~al.} 2018, \aap, 619,
  A147

\bibitem[{{Planck Collaboration} {et~al.}(2020){Planck Collaboration},
  {Aghanim}, {Akrami}, {Ashdown}, {Aumont}, {Baccigalupi}, {Ballardini},
  {Banday}, {Barreiro}, {Bartolo}, {Basak}, {Battye}, {Benabed}, {Bernard},
  {Bersanelli}, {Bielewicz}, {Bock}, {Bond}, {Borrill}, {Bouchet}, {Boulanger},
  {Bucher}, {Burigana}, {Butler}, {Calabrese}, {Cardoso}, {Carron},
  {Challinor}, {Chiang}, {Chluba}, {Colombo}, {Combet}, {Contreras}, {Crill},
  {Cuttaia}, {de Bernardis}, {de Zotti}, {Delabrouille}, {Delouis}, {Di
  Valentino}, {Diego}, {Dor{\'e}}, {Douspis}, {Ducout}, {Dupac}, {Dusini},
  {Efstathiou}, {Elsner}, {En{\ss}lin}, {Eriksen}, {Fantaye}, {Farhang},
  {Fergusson}, {Fernandez-Cobos}, {Finelli}, {Forastieri}, {Frailis},
  {Fraisse}, {Franceschi}, {Frolov}, {Galeotta}, {Galli}, {Ganga},
  {G{\'e}nova-Santos}, {Gerbino}, {Ghosh}, {Gonz{\'a}lez-Nuevo}, {G{\'o}rski},
  {Gratton}, {Gruppuso}, {Gudmundsson}, {Hamann}, {Handley}, {Hansen},
  {Herranz}, {Hildebrandt}, {Hivon}, {Huang}, {Jaffe}, {Jones}, {Karakci},
  {Keih{\"a}nen}, {Keskitalo}, {Kiiveri}, {Kim}, {Kisner}, {Knox},
  {Krachmalnicoff}, {Kunz}, {Kurki-Suonio}, {Lagache}, {Lamarre}, {Lasenby},
  {Lattanzi}, {Lawrence}, {Le Jeune}, {Lemos}, {Lesgourgues}, {Levrier},
  {Lewis}, {Liguori}, {Lilje}, {Lilley}, {Lindholm}, {L{\'o}pez-Caniego},
  {Lubin}, {Ma}, {Mac{\'\i}as-P{\'e}rez}, {Maggio}, {Maino}, {Mandolesi},
  {Mangilli}, {Marcos-Caballero}, {Maris}, {Martin}, {Martinelli},
  {Mart{\'\i}nez-Gonz{\'a}lez}, {Matarrese}, {Mauri}, {McEwen}, {Meinhold},
  {Melchiorri}, {Mennella}, {Migliaccio}, {Millea}, {Mitra},
  {Miville-Desch{\^e}nes}, {Molinari}, {Montier}, {Morgante}, {Moss}, {Natoli},
  {N{\o}rgaard-Nielsen}, {Pagano}, {Paoletti}, {Partridge}, {Patanchon},
  {Peiris}, {Perrotta}, {Pettorino}, {Piacentini}, {Polastri}, {Polenta},
  {Puget}, {Rachen}, {Reinecke}, {Remazeilles}, {Renzi}, {Rocha}, {Rosset},
  {Roudier}, {Rubi{\~n}o-Mart{\'\i}n}, {Ruiz-Granados}, {Salvati}, {Sandri},
  {Savelainen}, {Scott}, {Shellard}, {Sirignano}, {Sirri}, {Spencer},
  {Sunyaev}, {Suur-Uski}, {Tauber}, {Tavagnacco}, {Tenti}, {Toffolatti},
  {Tomasi}, {Trombetti}, {Valenziano}, {Valiviita}, {Van Tent}, {Vibert},
  {Vielva}, {Villa}, {Vittorio}, {Wandelt}, {Wehus}, {White}, {White},
  {Zacchei}, \& {Zonca}}]{Planck2020}
{Planck Collaboration}, {Aghanim}, N., {Akrami}, Y., {et~al.} 2020, \aap, 641,
  A6

\bibitem[{{Price} {et~al.}(2014){Price}, {Kriek}, {Brammer}, {Conroy},
  {F{\"o}rster Schreiber}, {Franx}, {Fumagalli}, {Lundgren}, {Momcheva},
  {Nelson}, {Skelton}, {van Dokkum}, {Whitaker}, \& {Wuyts}}]{Price2014}
{Price}, S.~H., {Kriek}, M., {Brammer}, G.~B., {et~al.} 2014, \apj, 788, 86

\bibitem[{{Reddy} {et~al.}(2022){Reddy}, {Topping}, {Shapley}, {Steidel},
  {Sanders}, {Du}, {Coil}, {Mobasher}, {Price}, \& {Shivaei}}]{Reddy2022}
{Reddy}, N.~A., {Topping}, M.~W., {Shapley}, A.~E., {et~al.} 2022, \apj, 926,
  31

\bibitem[{{Richard} {et~al.}(2021){Richard}, {Claeyssens}, {Lagattuta},
  {Guaita}, {Bauer}, {Pello}, {Carton}, {Bacon}, {Soucail}, {Lyon}, {Kneib},
  {Mahler}, {Cl{\'e}ment}, {Mercier}, {Variu}, {Tamone}, {Ebeling}, {Schmidt},
  {Nanayakkara}, {Maseda}, {Weilbacher}, {Bouch{\'e}}, {Bouwens}, {Wisotzki},
  {de la Vieuville}, {Martinez}, \& {Patr{\'\i}cio}}]{richard2021}
{Richard}, J., {Claeyssens}, A., {Lagattuta}, D., {et~al.} 2021, \aap, 646, A83

\bibitem[{{Roberts-Borsani} {et~al.}(2022{\natexlab{a}}){Roberts-Borsani},
  {Morishita}, {Treu}, {Brammer}, {Strait}, {Wang}, {Bradac}, {Acebron},
  {Bergamini}, {Boyett}, {Calabr{\'o}}, {Castellano}, {Fontana}, {Glazebrook},
  {Grillo}, {Henry}, {Jones}, {Malkan}, {Marchesini}, {Mascia}, {Mason},
  {Mercurio}, {Merlin}, {Nanayakkara}, {Pentericci}, {Rosati}, {Santini},
  {Scarlata}, {Trenti}, {Vanzella}, {Vulcani}, \&
  {Willott}}]{Roberts-Borsani2022b}
{Roberts-Borsani}, G., {Morishita}, T., {Treu}, T., {et~al.}
  2022{\natexlab{a}}, arXiv e-prints, arXiv:2207.11387

\bibitem[{{Roberts-Borsani} {et~al.}(2022{\natexlab{b}}){Roberts-Borsani},
  {Treu}, {Chen}, {Morishita}, {Vanzella}, {Zitrin}, {Bergamini}, {Castellano},
  {Fontana}, {Grillo}, {Kelly}, {Merlin}, {Paris}, {Rosati}, {Acebron},
  {Bonchi}, {Boyett}, {Bradac}, {Broadhurst}, {Calabro}, {Diego}, {Dressler},
  {Furtak}, {Filippenko}, {Glazebrook}, {Koekemoer}, {Leethochawalit},
  {Malkan}, {Mason}, {Mercurio}, {Metha}, {Nanayakkara}, {Pentericci},
  {Pierel}, {Rieck}, {Roy}, {Santini}, {Strait}, {Strausbaugh}, {Trenti},
  {Vulcani}, {Wang}, {Wang}, {Windhorst}, \& {Yang}}]{Roberts-Borsani2022c}
{Roberts-Borsani}, G., {Treu}, T., {Chen}, W., {et~al.} 2022{\natexlab{b}},
  arXiv e-prints, arXiv:2210.15639

\bibitem[{{Robertson} {et~al.}(2015){Robertson}, {Ellis}, {Furlanetto}, \&
  {Dunlop}}]{Robertson2015}
{Robertson}, B.~E., {Ellis}, R.~S., {Furlanetto}, S.~R., \& {Dunlop}, J.~S.
  2015, \apjl, 802, L19

\bibitem[{{Santini} {et~al.}(2022){Santini}, {Fontana}, {Castellano},
  {Leethochawalit}, {Trenti}, {Treu}, {Belfiori}, {Birrer}, {Bonchi}, {Merlin},
  {Mason}, {Morishita}, {Nonino}, {Paris}, {Polenta}, {Rosati}, {Yang},
  {Bradac}, {Calabr{\`o}}, {Dressler}, {Glazebrook}, {Marchesini}, {Mascia},
  {Nanayakkara}, {Pentericci}, {Roberts-Borsani}, {Scarlata}, {Vulcani}, \&
  {Wang}}]{Santini2022}
{Santini}, P., {Fontana}, A., {Castellano}, M., {et~al.} 2022, arXiv e-prints,
  arXiv:2207.11379

\bibitem[{{Saxena} {et~al.}(2022){Saxena}, {Cryer}, {Ellis}, {Pentericci},
  {Calabr{\`o}}, {Mascia}, {Saldana-Lopez}, {Schaerer}, {Katz}, {Llerena}, \&
  {Amor{\'\i}n}}]{Saxena2022}
{Saxena}, A., {Cryer}, E., {Ellis}, R.~S., {et~al.} 2022, \mnras, 517, 1098

\bibitem[{{Schaerer} {et~al.}(2022){Schaerer}, {Izotov}, {Worseck}, {Berg},
  {Chisholm}, {Jaskot}, {Nakajima}, {Ravindranath}, {Thuan}, \&
  {Verhamme}}]{Schaerer_2022}
{Schaerer}, D., {Izotov}, Y.~I., {Worseck}, G., {et~al.} 2022, \aap, 658, L11

\bibitem[{{Senchyna} {et~al.}(2022){Senchyna}, {Stark}, {Charlot}, {Plat},
  {Chevallard}, {Chen}, {Jones}, {Sanders}, {Rudie}, {Cooper}, \&
  {Bruzual}}]{Senchyna2022}
{Senchyna}, P., {Stark}, D.~P., {Charlot}, S., {et~al.} 2022, \apj, 930, 105

\bibitem[{{Sersic}(1968)}]{sersicpaper}
{Sersic}, J.~L. 1968, {Atlas de Galaxias Australes}

\bibitem[{{Sharma} {et~al.}(2016){Sharma}, {Theuns}, {Frenk}, {Bower}, {Crain},
  {Schaller}, \& {Schaye}}]{Sharma2016}
{Sharma}, M., {Theuns}, T., {Frenk}, C., {et~al.} 2016, \mnras
  [\eprint[arXiv]{1512.04537}]

\bibitem[{{Shibuya} {et~al.}(2015){Shibuya}, {Ouchi}, \&
  {Harikane}}]{Shibuya2015}
{Shibuya}, T., {Ouchi}, M., \& {Harikane}, Y. 2015, \apjs, 219, 15

\bibitem[{{Stark} {et~al.}(2010){Stark}, {Ellis}, {Chiu}, {Ouchi}, \&
  {Bunker}}]{Stark2010}
{Stark}, D.~P., {Ellis}, R.~S., {Chiu}, K., {Ouchi}, M., \& {Bunker}, A. 2010,
  \mnras, 408, 1628

\bibitem[{{Steidel} {et~al.}(2018){Steidel}, {Bogosavljevi{\'c}}, {Shapley},
  {Reddy}, {Rudie}, {Pettini}, {Trainor}, \& {Strom}}]{Steidel2018}
{Steidel}, C.~C., {Bogosavljevi{\'c}}, M., {Shapley}, A.~E., {et~al.} 2018,
  \apj, 869, 123

\bibitem[{{Thomas} {et~al.}(2013){Thomas}, {Steele}, {Maraston}, {Johansson},
  {Beifiori}, {Pforr}, {Str{\"o}mb{\"a}ck}, {Tremonti}, {Wake}, {Bizyaev},
  {Bolton}, {Brewington}, {Brownstein}, {Comparat}, {Kneib}, {Malanushenko},
  {Malanushenko}, {Oravetz}, {Pan}, {Parejko}, {Schneider}, {Shelden},
  {Simmons}, {Snedden}, {Tanaka}, {Weaver}, \& {Yan}}]{Thomas2013}
{Thomas}, D., {Steele}, O., {Maraston}, C., {et~al.} 2013, \mnras, 431, 1383

\bibitem[{{Trebitsch} {et~al.}(2022){Trebitsch}, {Dayal}, {Chisholm},
  {Finkelstein}, {Jaskot}, {Flury}, {Schaerer}, {Atek}, {Borthakur},
  {Ferguson}, {Fontanot}, {Giavalisco}, {Grazian}, {Hayes}, {Leclercq},
  {{\"O}stlin}, {Saldana-Lopez}, {Thuan}, {Wang}, {Worseck}, \&
  {Xu}}]{Trebitsch2022}
{Trebitsch}, M., {Dayal}, P., {Chisholm}, J., {et~al.} 2022, arXiv e-prints,
  arXiv:2212.06177

\bibitem[{{Treu} {et~al.}(2022){Treu}, {Roberts-Borsani}, {Bradac}, {Brammer},
  {Fontana}, {Henry}, {Mason}, {Morishita}, {Pentericci}, {Wang}, {Acebron},
  {Bagley}, {Bergamini}, {Belfiori}, {Bonchi}, {Boyett}, {Boutsia},
  {Calabr{\'o}}, {Caminha}, {Castellano}, {Dressler}, {Glazebrook}, {Grillo},
  {Jacobs}, {Jones}, {Kelly}, {Leethochawalit}, {Malkan}, {Marchesini},
  {Mascia}, {Mercurio}, {Merlin}, {Nanayakkara}, {Nonino}, {Paris},
  {Poggianti}, {Rosati}, {Santini}, {Scarlata}, {Shipley}, {Strait}, {Trenti},
  {Tubthong}, {Vanzella}, {Vulcani}, \& {Yang}}]{TreuGlass2022}
{Treu}, T., {Roberts-Borsani}, G., {Bradac}, M., {et~al.} 2022, \apj, 935, 110

\bibitem[{{Vanzella} {et~al.}(2019){Vanzella}, {Calura}, {Meneghetti},
  {Castellano}, {Caminha}, {Mercurio}, {Cupani}, {Rosati}, {Grillo}, {Gilli},
  {Mignoli}, {Fiorentino}, {Arcidiacono}, {Lombini}, \&
  {Cortecchia}}]{Vanzella2019}
{Vanzella}, E., {Calura}, F., {Meneghetti}, M., {et~al.} 2019, \mnras, 483,
  3618

\bibitem[{{Vanzella} {et~al.}(2016){Vanzella}, {de Barros}, {Vasei}, {Alavi},
  {Giavalisco}, {Siana}, {Grazian}, {Hasinger}, {Suh}, {Cappelluti}, {Vito},
  {Amorin}, {Balestra}, {Brusa}, {Calura}, {Castellano}, {Comastri}, {Fontana},
  {Gilli}, {Mignoli}, {Pentericci}, {Vignali}, \& {Zamorani}}]{Vanzella2016}
{Vanzella}, E., {de Barros}, S., {Vasei}, K., {et~al.} 2016, \apj, 825, 41

\bibitem[{{Verhamme} {et~al.}(2017){Verhamme}, {Orlitov{\'a}}, {Schaerer},
  {Izotov}, {Worseck}, {Thuan}, \& {Guseva}}]{Verhamme2017}
{Verhamme}, A., {Orlitov{\'a}}, I., {Schaerer}, D., {et~al.} 2017, \aap, 597,
  A13

\bibitem[{{Wang} {et~al.}(2019){Wang}, {Heckman}, {Leitherer}, {Alexandroff},
  {Borthakur}, \& {Overzier}}]{Wang2019}
{Wang}, B., {Heckman}, T.~M., {Leitherer}, C., {et~al.} 2019, \apj, 885, 57

\bibitem[{{Xu} {et~al.}(2022){Xu}, {Henry}, {Heckman}, {Chisholm}, {Worseck},
  {Gronke}, {Jaskot}, {McCandliss}, {Flury}, {Giavalisco}, {Ji}, {Amor{\'\i}n},
  {Berg}, {Borthakur}, {Bouche}, {Carr}, {Erb}, {Ferguson}, {Garel}, {Hayes},
  {Makan}, {Marques-Chaves}, {Rutkowski}, {{\"O}stlin}, {Rafelski},
  {Saldana-Lopez}, {Scarlata}, {Schaerer}, {Trebitsch}, {Tremonti}, {Verhamme},
  \& {Wang}}]{xu2022}
{Xu}, X., {Henry}, A., {Heckman}, T., {et~al.} 2022, \apj, 933, 202

\bibitem[{{Yamanaka} {et~al.}(2020){Yamanaka}, {Inoue}, {Yamada}, {Zackrisson},
  {Iwata}, {Micheva}, {Mawatari}, {Hashimoto}, \& {Kubo}}]{yamanaka2020}
{Yamanaka}, S., {Inoue}, A.~K., {Yamada}, T., {et~al.} 2020, \mnras, 498, 3095

\bibitem[{Yang {et~al.}(2020)Yang, Birrer, \& Treu}]{Yang_2020}
Yang, L., Birrer, S., \& Treu, T. 2020, Monthly Notices of the Royal
  Astronomical Society, 496, 2648

\bibitem[{{Yang} {et~al.}(2022){Yang}, {Leethochawalit}, {Treu},
  {Roberts-Borsani}, {Brada{\v{c}}}, {Birrer}, {Castellano}, {Merlin},
  {Fontana}, {Amorin}, \& {Trenti}}]{Yang2022}
{Yang}, L., {Leethochawalit}, N., {Treu}, T., {et~al.} 2022, \mnras, 514, 1148

\bibitem[{{Zackrisson} {et~al.}(2017){Zackrisson}, {Binggeli}, {Finlator},
  {Gnedin}, {Paardekooper}, {Shimizu}, {Inoue}, {Jensen}, {Micheva},
  {Khochfar}, \& {Dalla Vecchia}}]{Zackrisson2017}
{Zackrisson}, E., {Binggeli}, C., {Finlator}, K., {et~al.} 2017, \apj, 836, 78

\bibitem[{{Zackrisson} {et~al.}(2013){Zackrisson}, {Inoue}, \&
  {Jensen}}]{Zackrisson2013}
{Zackrisson}, E., {Inoue}, A.~K., \& {Jensen}, H. 2013, \apj, 777, 39

\end{thebibliography}

\end{document}